# Semantic annotation for computational pathology: Multidisciplinary experience and best practice recommendations

Noorul Wahab[1], Islam M Miligy[2,3], Katherine Dodd[4], Harvir Sahota[4], Michael Toss[2], Wenqi Lu[1], Mostafa Jahanifar[1], Mohsin Bilal[1], Simon Graham[1], Young Park[1], Giorgos Hadjigeorghiou[1], Abhir Bhalerao[1], Ayat Lashen[2], Asmaa Ibrahim[2], Ayaka Katayama[5], Henry O Ebili[2], Matthew Parkin[2], Tom Sorell[6], Shan E Ahmed Raza[1], Emily Hero[4,7], Hesham Eldaly[4], Yee Wah Tsang[4], Kishore Gopalakrishnan[4], David Snead[4], Emad Rakha[2], Nasir Rajpoot[1], Fayyaz Minhas[1]

[1]Tissue Image Analytics Centre, University of Warwick, Coventry, UK
[2]University of Nottingham, Nottingham, UK
[3]Department of Pathology, Faculty of Medicine, Menoufia University, Egypt
[4]University Hospital Coventry and Warwickshire, Coventry, UK
[5]Gunma University, Graduate School of Medicine, Maebashi, Japan
[6]Department of Politics and International Studies, University of Warwick, Coventry UK
[7]University Hospitals Leicester, UK

Corresponding author: Noorul Wahab
**Postal address:** Tissue Image Analytics (TIA) Centre, Department of Computer Science, University of Warwick, Coventry, UK, CV4 7AL
**Telephone number:** +44 (24) 7657 3795
**Email:** noorul.wahab@warwick.ac.uk

## Abstract

Recent advances in whole slide imaging (WSI) technology have led to the development of a myriad of computer vision and artificial intelligence (AI) based diagnostic, prognostic, and predictive algorithms. Computational Pathology (CPath) offers an integrated solution to utilize information embedded in pathology WSIs beyond what we obtain through visual assessment. For automated analysis of WSIs and validation of machine learning (ML) models, annotations at the slide, tissue and cellular levels are required. The annotation of important visual constructs in pathology images is an important component of CPath projects. Improper annotations can result in algorithms which are hard to interpret and can potentially produce inaccurate and inconsistent results. Despite the crucial role of annotations in CPath projects, there are no well-defined guidelines or best practices on how annotations should be carried out. In this paper, we address this shortcoming by presenting the experience and best practices acquired during the execution of a large-scale annotation exercise involving a multidisciplinary team of pathologists, ML experts and researchers as part of the **Path**ology image data **L**ake for **A**nalytics, **K**nowledge and **E**ducation (PathLAKE) consortium. We present a real-world case study along with examples of different types of annotations, diagnostic algorithm, annotation data dictionary and annotation constructs. The analyses reported in this work highlight best practice recommendations that can be used as annotation guidelines over the lifecycle of a CPath project.

# 1. Introduction

Recent developments in imaging technology, digitization of glass slides and Artificial Intelligence (AI) have spurred an ongoing revolution in clinical histopathology workflows and enabled automated analysis of digital pathology whole slide images (WSIs). This is evidenced by growth in the commercial and government investment in Computational Pathology (CPath) as well as the rapid rise in the number of scientific publications in this field (1,2). In the United Kingdom, the **Path**ology image data **L**ake for **A**nalytics, **K**nowledge and **E**ducation (PathLAKE) consortium has been supported by £15 million fund to create a unique data resource of pathology images (a 'data lake') and develop AI technologies for cancer diagnosis and personalized treatment for routine clinical practice. The PathLAKE consortium brings together academic researchers, the UK National Health Service (NHS) hospital Trusts and leading AI and software business developers to create a national centre of excellence in CPath that will be linked to multiple NHS pathology laboratories. PathLAKE is poised to serve as a technology demonstrator at the national scale for implementation of CPath workflows. Through its exemplar projects such as the Breast Cancer (BraCe) prognostication, PathLAKE aims to pioneer routine clinical deployment of AI algorithms. Similar large-scale CPath projects are underway elsewhere, such as the BIGPICTURE initiative (3).

CPath algorithms utilize the fact that there is fundamental information of diagnostic and prognostic benefit embedded in WSIs (4). The typical lifecycle of a histological image analysis project in CPath is shown in Fig. 1. Digitized tissue slides may be viewed online for remote consultation and can be processed by digital image processing and machine learning (ML) algorithms for the development of diagnostic and prognostic tools (5). However, before the sample can be analyzed digitally, the tissue must be sampled, processed, sectioned, subjected to antigen retrieval, and stained using chemical, immunohistochemical (IHC), or immunofluorescent (IF) techniques and then digitized with a slide scanner resulting in very large (100,000 × 100,000 pixels resolution) whole slide images (WSIs). Each of these steps introduces preanalytical or analytical incongruencies which, on top of tumor heterogeneity and biological variability, makes it challenging to obtain consistent input data for computational analysis. The ability to mine "sub-visual" image features in WSIs that may not be discernable to a pathologist, such as cell-to-cell interactions offers the opportunity for better quantitative modelling of disease morphology and hence the possibility of improved prediction of disease aggressiveness and patient outcome (6).

Using ML in CPath can help overcome major challenges in clinical practice such as data or material sharing, subjectivity (7), non-reproducibility due to inter- and intra- observer disagreement (8), inefficiency in the form of delayed and inaccurate diagnoses (9) and integration with genomic or multidomain data (10,11). With an ever-increasing number of patients, a global decrease in doctors enrolling to the pathology specialty and an aging workforce (12,13), time and cost efficiency in clinical pathology is of utmost importance. AI methods can help reduce diagnosis times by automating the process or assisting pathologists in various diagnostic steps such as detecting regions of interest, highlighting different tissue types, counting cells, visualizing clinically important structures and features, lymph node assessment for metastasis, and analyzing a WSI for grading (14).

The quality of annotation data has a direct impact on the efficacy of the resulting ML-based CPath solution. A sufficient amount of well-labelled/annotated/curated data is required to train ML models (15,16). With approaches such as self-supervised learning, weak supervision, domain adaptation and transfer learning, there has been significant progress in ML on using a small amount of annotated data for training algorithms that are robust and generalize well to unseen WSIs (17–23). However, validation of even these label-efficient algorithms and root-cause

Recommendations on semantic annotation for computational pathology analysis of failure cases of these algorithms still requires annotations. In addition to the requirement of annotations for training ML techniques, clinical deployment of these methods also warrants using well-annotated samples for strong validation to ensure robustness and interpretability of "black-box" or "grey-box" AI models (24). Generating these annotations is a labor-intensive process because of the large volumes of data involved. Crowdsourcing may be cheaper and quicker but has the potential of introducing inconsistency, inaccuracies, and the difficulty of maintaining quality control, as well as ethical issues of sharing clinically sensitive data (15).

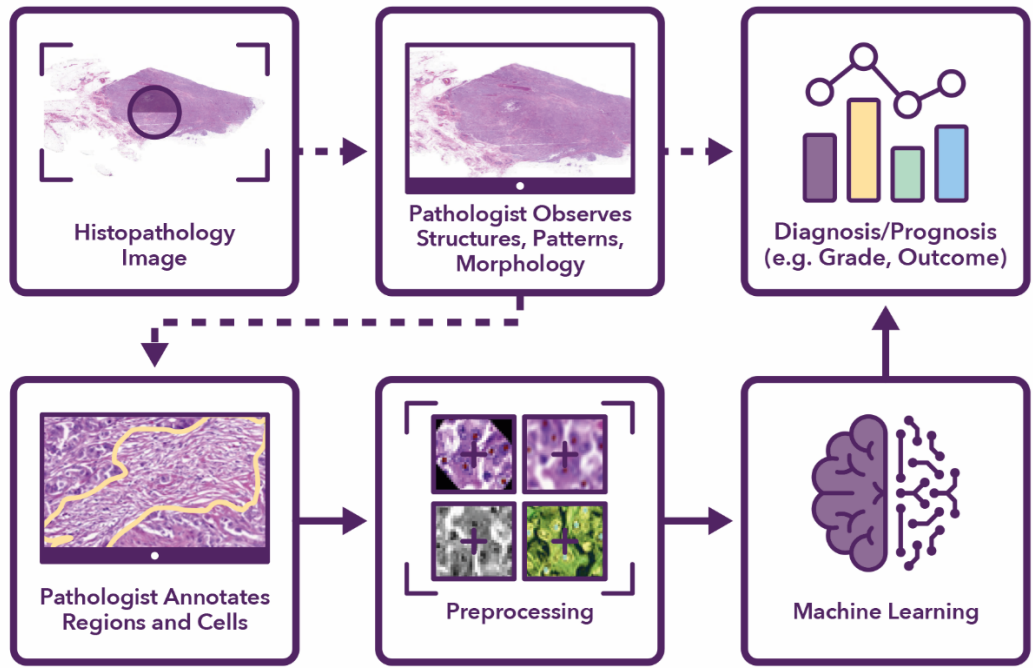

**Fig. 1** Manual vs automated process of histopathology image-based diagnosis/prognosis. The dotted arrows show the manual process whereas the solid arrows show the steps involved in automating the process.

There is no single methodology for annotating different structures and no agreed data format for storing annotations. The diversity of CPath solutions in terms of their objectives and diverse tissue types add to these challenges. Therefore, without any guidelines on how these annotations should be collected and used, there may be a significant repetition of effort across different CPath projects in quality assurance for making, managing, and using annotations. The development of such guidelines could help streamline the annotation process by providing definitions and standard operating procedures (SOPs) which can guide an annotation team specifying in detail when to annotate what, and how to annotate histological structures for effective downstream AI use. Such guidelines could enable multi-center projects such as PathLAKE to collaborate effectively across different centers, assist with training new staff and educating future pathologists. In addition, these guidelines may help ensure the reproducibility of results and the robust validation and testing of future ML algorithms, which is of great value when seeking regulatory approval for a new computational method in pathology.

To the best of our knowledge, there are no existing guidelines that specifically address annotation exercises for largescale for CPath projects involving multidisciplinary teams. A related work (25) describes annotation of regions in 3D images and compares different annotation software. It also includes an automatic tool for 'assigning' cases to annotators and then comparing the confidence in the annotations. Though Grünberg *et al.* (25) describe some aspects of 3D annotations, no detailed guidelines are provided for 2D images. The Digital Pathology Association has defined terms and concepts of CPath with a focus on histology images. They recommend best practices for CPath workflow implementation such as training data acquisition, quality assessment, and ethical, regulatory and security concerns. Separately, Willemink *et al.* (26) have proposed steps involved in preparing medical imaging data for ML algorithm development and have described the limitations of data curation.

To address the above challenges in annotations for CPath projects, we propose a comprehensive set of annotation guidelines in this paper based on our practical experience and recent involvement with PathLAKE exemplar projects. We hope this will pave the way for the interoperability of annotation protocols and the improved generalizability of algorithms via multicenter validation.

## 2. Materials and Methods

In this section, we discuss our proposed workflow for semantic annotation of pathology images for CPath projects.

## Proposed annotation workflow

The proposed annotation workflow is illustrated in Fig. 2 and each step is further discussed in the following subsections. The proposed workflow begins with framing the underlying ML problem by documentation of project objectives and the diagnostic or prognostic process used in clinical practice in the form of an algorithm. This documentation leads to the definition of various 'annotation constructs' required for ML algorithm development. The specification document of these constructs forms a reference 'data dictionary' that leads to the definition of 'levels', 'degrees' and 'phases of annotations' to capture anatomical constructs essential in the context of the specific goals of a project at different stages while keeping the project timeline, annotators' time and annotation budget in consideration. We also propose a stringent quality review to ensure following all the steps properly to yield standardized annotations.

## 2.1 Definition of project objectives

The annotation process will be guided mainly by the objectives of the project. Therefore, we propose defining project objectives as the first step in the annotation workflow which will help to align the annotation protocol with these objectives. For example, if the main objective of a CPath project is to automate the process of grading breast cancer (BCa) in WSIs, then the different structures relevant to the grading process, such as, tubules, tumor cell morphology, and mitotic figures will be part of the annotation protocol. These features will guide the annotation process by allowing the definition of the diagnostic algorithm and the subsequent annotations required to enable an ML model to assign a grade to an input image. Depending on a project's objective, another important aspect is the estimation of sample size (27,28). A large sample size may be required when developing a prognostic algorithm where the objective is to measure very minute differences between groups.

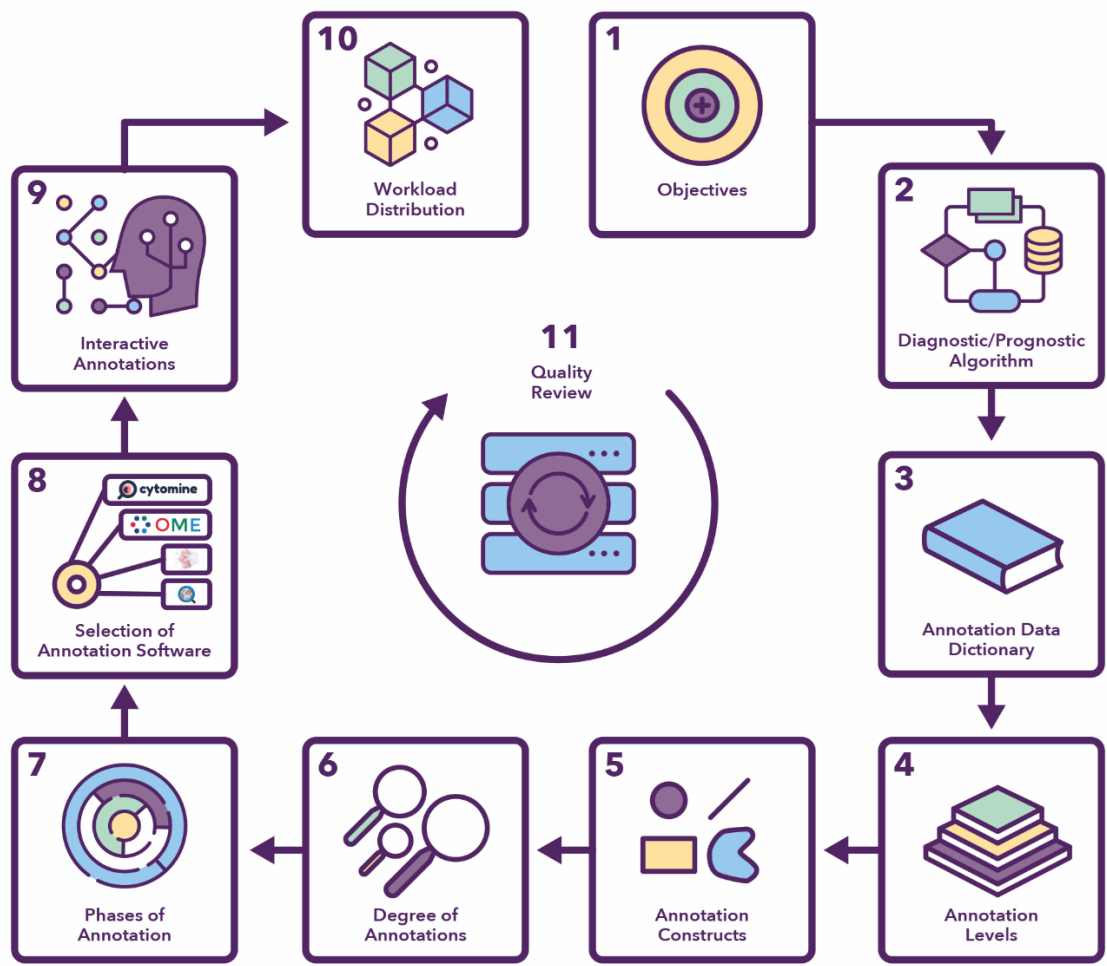

**Fig. 2** Proposed annotation workflow for a CPath project

## 2.2 Development of clinical diagnostic and/or prognostic algorithms

In order to identify relevant clinical and diagnostic constructs for the quality annotations downstream ML solutions, we propose developing a clinical diagnostic/prognostic algorithm as the second step in the CPath project. Such algorithms enumerate steps that pathologists would perform for the routine diagnosis or prognosis of the disease or outcome of interest. The development of clear and accurate clinical algorithm will guide the rest of the annotation workflow and ensure a clear understating of the significant aspects of the problem by the multidisciplinary project team. With our example of BCa grading, the diagnostic algorithm would model the Nottingham Scoring System, which assigns scores for tubule formation, nuclear pleomorphism, and mitotic count. Once such an algorithm is in place, it clarifies what to annotate for training an ML model.

## 2.3 Development of an annotation data dictionary

We propose the development of a *data dictionary* for every annotation project which is a standard reference document throughout the life cycle of the project. Realization of the diagnostic algorithm forms the basis of the data dictionary and defines different annotation constructs. The dictionary can serve multiple purposes. It facilitates agreement on regions and cells by clearly defined terms. It adds value to annotations, acts as metadata for annotations, and enables long term use of annotations. It avoids the drift of definitions over the project lifetime. It also acts as a communication tool between pathologists, ML experts, and other collaborators. The data dictionary may include information to answer common project-related questions, for example: What needs to be annotated? What is the diagnostic/prognostic value of each annotation type? What order to follow for annotations? How much to annotate (e.g., exhaustive, non-exhaustive)? Clear examples of typical diagnostic cases with illustrative images facilitate the task of training new project staff.

A part of the PathLAKE data dictionary for our BC exemplar project is shown in supplement Fig. S1 where different bounding boxes are defined for region-level and cell-level annotations. Supplement Tables (S1, S2, S3, S4) define bounding boxes, region types, cell types, and annotation styles for regions, respectively. Annotation style files (in JSON format in this case) are defined according to the data dictionary and are then used to create annotations using the annotation software.

## 2.4 Defining annotation levels

For achieving the aims and objectives of an ML project, annotations should be marked at different levels of detail. For example, keeping the case/slide-level annotations at the first level can make the computational analysis efficient since it is less time-consuming that marking annotation constructs at region and cell level. A more detailed level analysis, which may use more explainable features, will require further detailed annotations of the individual WSI. Fig. 3 shows the four levels of annotations.

### 2.4.1 Case-level annotations

Depending on the underlying task, a single label can be assigned to a case and its representative slides or to each slide separately. The label can be binary (such as benign vs malignant) or multi-class (e.g., grade 1, 2, 3). These labels represent the overall diagnosis or prognosis. For example, if the case is grade 1, 2 or 3, or if it belongs to a high or low risk group. Such annotations can be used by ML models which do not require detailed annotations, known as weakly-supervised ML models. Deep convolutional neural network can be used as weakly-supervised classifier to extract useful features from the raw image data. For this purpose, the image is usually divided into multiple tiles for processing and aggregating the results to make a case-level prediction (29). In such a scenario, multiple instance learning (MIL) or other methods of aggregation can be used (30–32).

### 2.4.2 Region-level annotations

Region level annotations include different regions in the WSI which can be of any diagnostic or prognostic value for a project at hand, for example, tumor, stromal and ductal carcinoma in-situ (DCIS) regions in breast cancer. Supplement Fig. S2 shows examples of region-level annotations. For flexibility in the degree of granularity in the downstream analysis, a large number of region types can be allowed for region-level annotations initially and merged later, if necessary. For example, invasive tumor region in BCa images can be differentiated into tubular, non-tubular, and mucinous regions. Region-level annotations can be used for training region segmentation models as well as restricting the detailed analysis to a specific area, for example, calculating the percentage of positive invasive tumor cells out of all invasive tumor cells for nuclear staining to Ki-67. Similarly, stromal region segmentation is required to obtain the percentage of tumor-infiltrating lymphocytes (TILs). Depending on ML needs, patches of different sizes can be extracted from the annotated regions for training/validating ML models. Region-level annotations are usually carried out by drawing polygons around the areas, but bounding boxes can similarly be used.

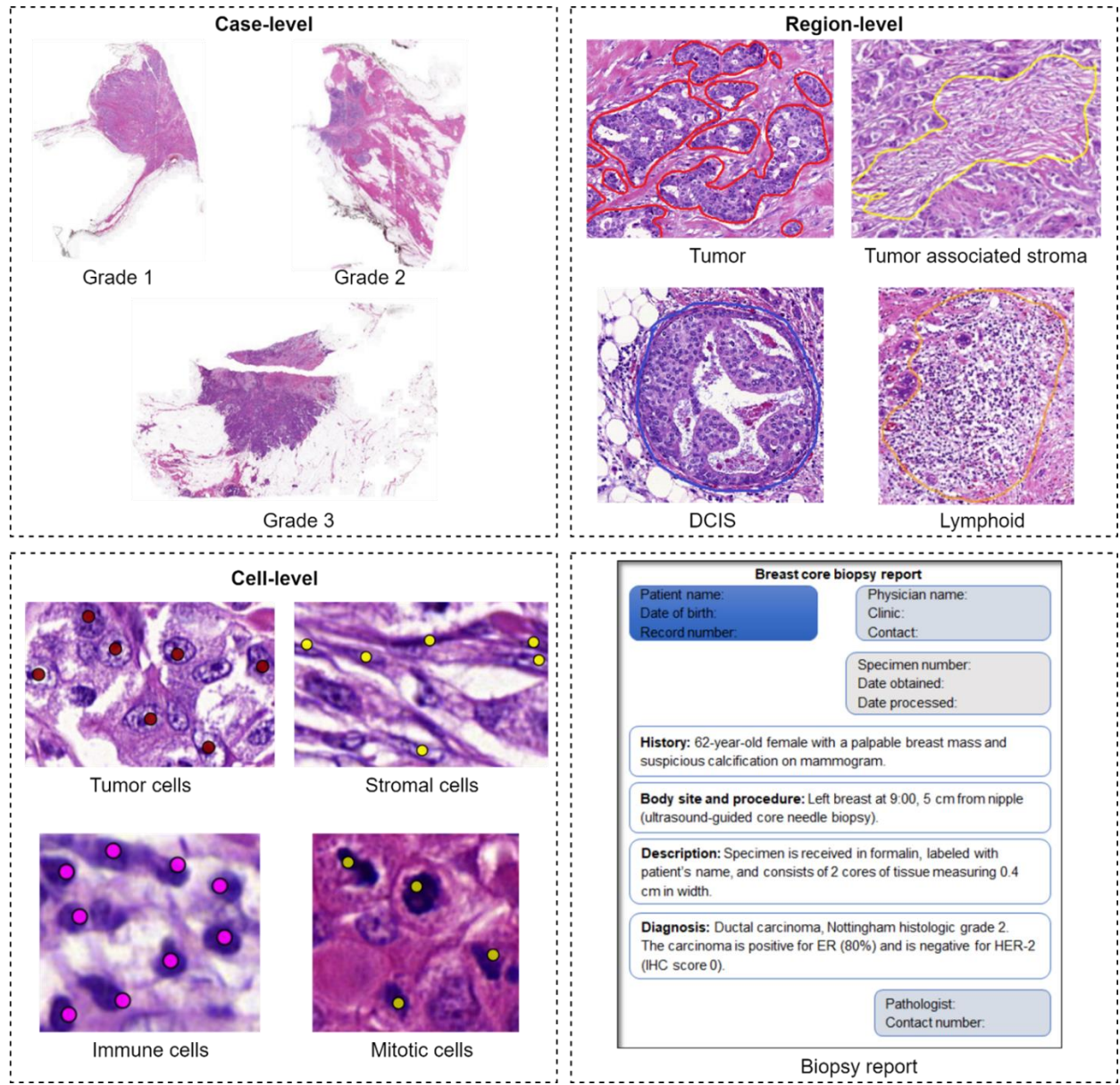

**Fig. 3** Four levels of annotations

### 2.4.3 Cell-level annotations

Cell-level annotations are the more detailed annotations in which different cell types are either annotated by a point/dotting (for cell detection and classification) or by marking the boundary of nuclei in a free-hand manner if cell segmentation is necessary as part of the downstream analysis. Supplement Fig. S3 shows both point and free-hand (polygon) annotations of different cells. Various features are extracted from cells and their environment for downstream diagnostic or prognostic purposes. These features include morphology, cell counts, cell-to-cell ratios, cell graphs, and cellular colocalization.

### 2.4.4 Descriptive and multi-modal annotations

Annotations in the form of textual description such as a pathology report or text annotation of an area of interest on the WSI can also be recorded. For example, BCa pathology report provides clinically relevant information including diagnosis, invasive tumor size, lymph node status and potential targets for treatment (e.g., hormone receptor status). A natural language processing system can be used to automate the process of extracting useful information from the descriptive annotations but as the descriptions may vary from laboratory to laboratory, it might be challenging to achieve high accuracy (33).

In addition, with a growing interest in the integration of histopathology with genomics and transcriptomics, associated data can be collected from other sources to augment imaging information to advance patient care (11,34). Genomic, epigenomic, transcriptomics, proteomic and imaging data can be obtained from different sources such as The Cancer Genome Atlas (TCGA) and Clinical Proteomic Tumor Analysis Consortium (CPTAC).

## 2.5 Defining annotation constructs

Different annotation shapes can be used for different types and levels of annotations. The main annotation constructs include bounding box, point/circle, polygon, line, and text (supplement Section S1.4 and Fig. S4). Different line-style, linewidth, line-color, fill-color etc. can be defined in the annotation protocol to use the same constructs for further categorization. For example, a green-colored polygon might be used to represent a normal area whereas a red-colored polygon may indicate an area of tumor (supplement Table S4).

## 2.6 Degree of annotation

Annotations of CPath WSIs can be performed at varying degrees of exhaustiveness. In exhaustive annotations, all the features that exist inside a *containing construct* (such as a freehand polygon, bounding box or even the entire WSI) are annotated. Such annotations make the evaluation of an ML model easier. If every cell inside a box is annotated, then it is possible to compare the predictions of the model with the ground-truth. The required number of cell-level and region-level boxes should be defined so that each WSI could be checked for completeness of the annotations. In non-exhaustive annotations, regions, and cells of interest in different areas of the image are marked in a non-exhaustive manner (supplement Fig. S5). With such annotations some structures may be omitted, even if they are defined in a protocol. They are beneficial where rare objects e.g., mitoses or ROIs, are being annotated. Non-exhaustive annotations are good for obtaining diverse annotations as the annotator is not restricted to a particular image region. However, for non-exhaustive annotations, if the ML model predicts some cells or regions in the image(s) that were not annotated, the training and evaluation processes of the ML model may need to be modified to take account of non-annotated area(s) in the image (7).

## 2.7 Phases of annotations

We recommend that annotations be carried out in a phased manner where each phase focuses on a particular level of annotations (supplement Fig. S6). A CPath project team may find a pilot phase of annotation beneficial to identify issues regarding the usability of annotation tools, understanding of the data dictionary and level of agreement between constructs, regions, and cells. This phase will help in avoiding many of the problems that can arise during the subsequent phases of the project. The pilot phase will help train the annotation team and familiarize them with new constructs and terms defined in the data dictionary. In the first phase, the slides are assigned a case- or slide-level label. Case-level annotations can be used in a weakly supervised manner for building ML models. This will also help in identifying important regions and cells for the problem at hand and will guide the process of annotation prioritization. Depending on the nature of problem, the second phase can either be region-level or cell-level annotations.

## 2.8 Selection of annotation software

Annotation of a WSI is a detailed and time-consuming process for pathologists. It is important therefore to use software that is user-friendly and easily accessible. The following factors should be considered when selecting an annotation tool: Does it support all the annotation constructs defined for the project? Is it easy enough to be used by the annotators or does it require extensive training? Is it web-based or desktop-based? Does it support different image types/formats? Does it have a workflow module, including the ability to configure a data dictionary and annotation style to all annotators for a CPath project (Supplement Table S4)? How does it store the annotations and the related meta-data? How secure is the system? Does it support adding annotation completeness criteria? Can it search by case or annotation type? Can it be upgraded securely to add extra functionalities without losing the work that has been already done before on it?

For a small project, a desktop-based annotation tool will work but for larger projects a web-based interface of the annotation software will be essential. A web-based annotation tool provides the flexibility to annotate images from any location. There are several open-source tools available for annotating histopathology images (supplement Table S5). It would also be useful to describe the steps of doing the annotations using the selected software and defined data dictionary (supplement Fig. S7).

## 2.9 Interactive and active annotations

When the annotation budget is small or there is a limited availability of annotation experts, interactive annotation and active learning can be used to speed up exhaustive annotations. In interactive annotation the user reviews the output of the annotation model and provides feedback to improve model's performance. Active learning works in an iterative manner where the annotation model asks the user (teacher) for samples from an unannotated dataset such that the performance of the model improves. Supplement section S1.5 mentions some advantages of interactive over fully automated approaches. Supplement Fig. S8 shows how an interactive segmentation tool (NuClick (22)) can be used to generate nuclei and gland masks from point and squiggle annotations.

Similarly, active learning-based annotations can also help to reduce the need for manual annotations and speed up the of annotation process (35). In this setting, an ML algorithm (for example, a region segmentation model) utilizes a limited annotation set for training. It then produces annotations as output which are then confirmed or modified by a pathologist. The augmented annotations are reused for training the model and the process is repeated until a desired level of accuracy is achieved.

## 2.10 Workload distribution

It is not trivial to accurately estimate workload distribution because of the complex nature of histopathology image annotation, involving different levels (cases, regions, cells, descriptive reports), details (exhaustiveness, concordance), and pathologists' experience, clinical time constraints, and daily work commitments (36). A better distribution of workload can be arrived at by listing the number of cases to annotate, the number and types of annotations per phase, timeframe, and the number of available pathologists. A pilot phase or initial analysis of the annotation might be helpful in the workload estimation. Similarly, an annotation tool facilitating the automatic assignment of annotation tasks cam ease assigning cases to pathologists for annotation.

## 2.11 Quality review

Quantitative analysis relies on the quality of the WSIs which in turn depends on the quality of tissue sectioning, staining and scanning. Supplement section 1.6 describes how these steps can affect the annotations and hence the ML based analysis.

### 2.11.1 Quality control of images

The staining and scanning quality of images is important for good annotations and hence better ML models. Only images passing Image quality control (QC) (37) should be included for further processing and analysis. In PathLAKE at Warwick, we have used a reproducible and automated image quality analysis pipeline (ImageQC) for precisely localized artefacts to identify slides that need to be re-scanned or regions that should be avoided during computational analysis. ImageQC workflow includes five steps: i) Pen markings detection, ii) Coverslip edge detection, iii) Tissue detection, iv) Blurry region identification, and v) Morphological operations (supplement Fig. S9). Briefly, each whole slide image goes through pen-markings and coverslip edge detections to remove shadow areas and pen markings, followed by the tissue region detection in the non-pen-markings regions. Then blur detection method (38) is used to detect the blurriness in the detected tissue region. The blur detection method generates a blurred version of the input image and then estimates blurriness based on the variation between neighboring pixels in the input versus blurred images. Based on empirical evidence a slides with more than 60% blurry regions will be excluded for further analysis and returned for rescanning. After getting the artefact free regions using the first four steps, we then perform several morphological operations to generate a single tissue mask which is ready for further annotation and computational analysis. These operations cover removing small disconnected foreground objects, and filling small background holes in the mask.

### 2.11.2 Annotation quality

During the whole process of annotation, annotation quality should be regularly reviewed. This can also help annotators identify their errors and improve the annotation quality over time. There should be a ticket-and-trace mechanism included in the annotation workflow for resolving any annotation errors. All issues and their resolution should be documented in a shared annotation quality log for error tracking and future reference.

# Recommendations on semantic annotation for computational pathology

### 2.11.2.1 Quality control metrics for annotation

Depending on the types of annotations different QC steps can be defined. Fig. 4 shows generic steps involved in the automatic QC and analysis of the annotations. For a detailed QC of annotations, we propose four metrics to measure completeness, exhaustiveness, diversity, and agreement of annotations (Table 1).

For the annotation data dictionary, 'completeness' criterion ensures that the annotations for an image are complete. This includes a check as to whether the required number of cell-level and region-level boxes have been annotated by the required number of annotators. The exhaustiveness criteria make sure that all the structures (regions, cells, etc.) in a region-box are annotated as much as possible. To obtain a sufficient percentage of annotated regions, some basic tissue segmentation/thresholding is required so that the non-tissue area is discarded in the calculations. The exhaustiveness may not often reach 100% because for some regions it may be impossible to discern precise start and end points or are subjective in whether they comply precisely with the definition. The annotators may sometime omit an object because of its lack of relevance to the annotation protocol. Likewise, where the regions are annotated by multiple annotators, exhaustiveness will be calculated for regions on which all the annotators agree. Then, based on some initial annotations, a threshold can be defined to identify cases not satisfying the exhaustiveness criterion. Similarly, the agreement criterion measures the agreement between multiple annotators. To measure such inter-annotator agreement, the Jaccard index (eq. 1) which calculates the ratio of the intersection of the annotated areas to the union of the annotated areas can be used. In eq. 1, if A and B represent the regions annotated by two different annotators, $|A \cap B|$ is the number of elements (pixels) where annotations of two annotators intersect and $|A \cup B|$ would be the union of both A and B annotations. The value of the Jaccard index range between 0 (no agreement at all) and 1 (100% agreement). Other similar metric such as Dice index could also be used instead of Jaccard index.

$$J(A, B) = \frac{|A \cap B|}{|A \cup B|}$$
<div align="right">eq. 1</div>

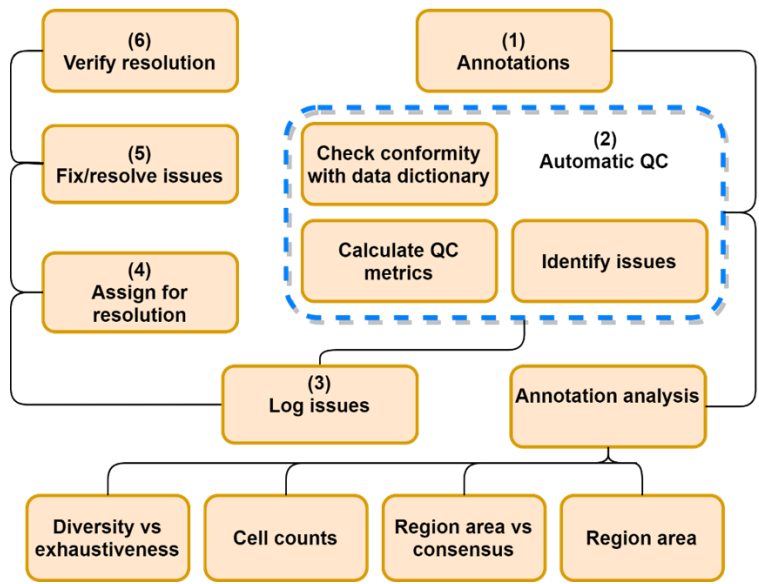

**Fig. 4** Annotation Quality Control steps

**Table 1** Proposed annotation quality control metrics

| Matric name | Purpose | Unit |
|---|---|---|
| Completeness | Are the annotations complete according to the defined protocol? | Yes/No |
| Exhaustiveness | What percentage of tissue is annotated in the defined box(es)? | Percentage area |
| Diversity | How many types of regions are annotated? | 1 to number of defined types in the protocol. |
| Agreement | How much the annotators agree? | Jaccard index |

## 2.11.2.2 Automatic quality control of annotations

Manual review of all the annotations is a time-consuming task. For large CPath projects, an automatic QC pipeline is required to identify problematic annotations which can then be reviewed by pathologists. As a first step in automatic QC, annotations for all the images are checked for completeness. All cases that satisfy the completeness criterion are then passed through the other three stages while the remaining cases which have failed the completeness check are logged for review by the annotation team. To make the process of logging and rectifying an annotation easier, a ticket-based tracking system can be used. After the automatic QC, problematic annotations are logged in the system with a unique annotation ID, WSI ID, logged date, and description of the issue. A nominated QC member from the annotation team then triages the issues and assigns to another annotator to resolve the issue. Once resolved, the issue is updated with the resolution details and a date stamp in the log database.

## 2.11.2.3 Pathologist's review of annotations

We recommend a regular review of the annotations by pathologists so that any major issues in the annotation process can be identified and rectified in a timely manner. The review process can also cover any issues brought up during the QC. Relatively minor issues can include improper use of styles, annotation constructs etc., and these may be resolved by the individual pathologist to whom those annotations were assigned. Other important issues such as disagreement on features should be discussed for resolution at the periodic review meetings.

## 2.11.2.4 Annotation interoperability

For annotations to be interoperable with other software tools, there should a proper schema defined for all the styles and structures so that there is minimal overhead for translation for use with other systems. Annotation schema help standardize the annotations and smooth the conversion process if the project involves annotations from multiple centers using different annotation software.

# 3. Results (Applied example)

In this section we present the results of applying the guidelines discussed above for the Breast Cancer (BraCe) project under the PathLAKE consortium so that these can be used as a guide for future projects. The cases and the corresponding annotations for BraCe were collected from Nottingham City Hospital UK where seven pathologists (IM, MT, AL, AI, AK, HOE, and MP) were involved in the annotation process.

## 3.1 Project objectives

The objectives of the BraCe project were clearly defined in a project document and included in the data dictionary. The main objective of the BraCe project is "Automatic analysis of breast cancer whole slide images for grading and prognosis".

## 3.2 Diagnostic/prognostic algorithm

In line with the project objective, a clear and detailed clinical diagnostic algorithm for BCa grading was specified in Fig. 5. The algorithm provides a holistic view of the different steps involved in the assessment process. In BCa grading three main features (tubule formation, nuclear pleomorphism, and mitotic count) are identified to calculate and assign a grade to each case.

## 3.3 Annotation data dictionary

The main parts of the BraCe annotation data dictionary are provided in supplement Fig. S1 and Tables S1-S4. Supplement Fig. S1 shows the different types of bounding boxes and describes their purpose. Supplement Table S1 further describes each bounding box, for example, Box_All_Cell_20x is required for cell-level consensus annotations by multiple pathologists. Supplement Tables S2 and S3 describe each region and cell types, respectively that are required to be annotated. Finally, supplement Table S4 lists all the styles (name, style name, annotation shape, line color, and the Red Green Blue (RGB) values for the line colors) to differentiate them visually more easily as well as for ML purposes.

## 3.4 Annotation levels

Annotations were conducted at three levels (case-level, region-level, and cell-level). Table 2 lists the details of annotations made at different levels. Note that the case-level annotations differ for the different stains because of some cases dropped during QC. Supplement (Fig. S2) shows some sample region-level annotations where different types of regions are annotated in different colors. Similarly, supplement Fig. S3 shows both point and free-hand polygon annotations of different types of cells. Seventy cases were used for consensus as well as individual annotations whereas the rest were used for annotations by individual annotators only. The consensus annotations were completed by two annotators. Supplement Tables S6-S11 list details of the numbers and areas of region-level and counts of cell-level annotations for both H&E and IHC stained slides. Region-level annotations were done at 5× magnification whereas cell-level annotations were done at 20× magnification. For each case two bounding boxes were drawn for regions and two bounding boxes for cells.

Annotations were completed in different batches of seven sets whereas each set comprise 70-80 cases and each has 3 images (H&E, PR and Ki67) to allow modular annotation and to analyze and compare them properly. Supplement Fig. S10A shows a timeline of the overall region- and cell-level annotations whereas Figs. S10B and S10C present details of the different types of region-level and cell-level annotations, respectively.

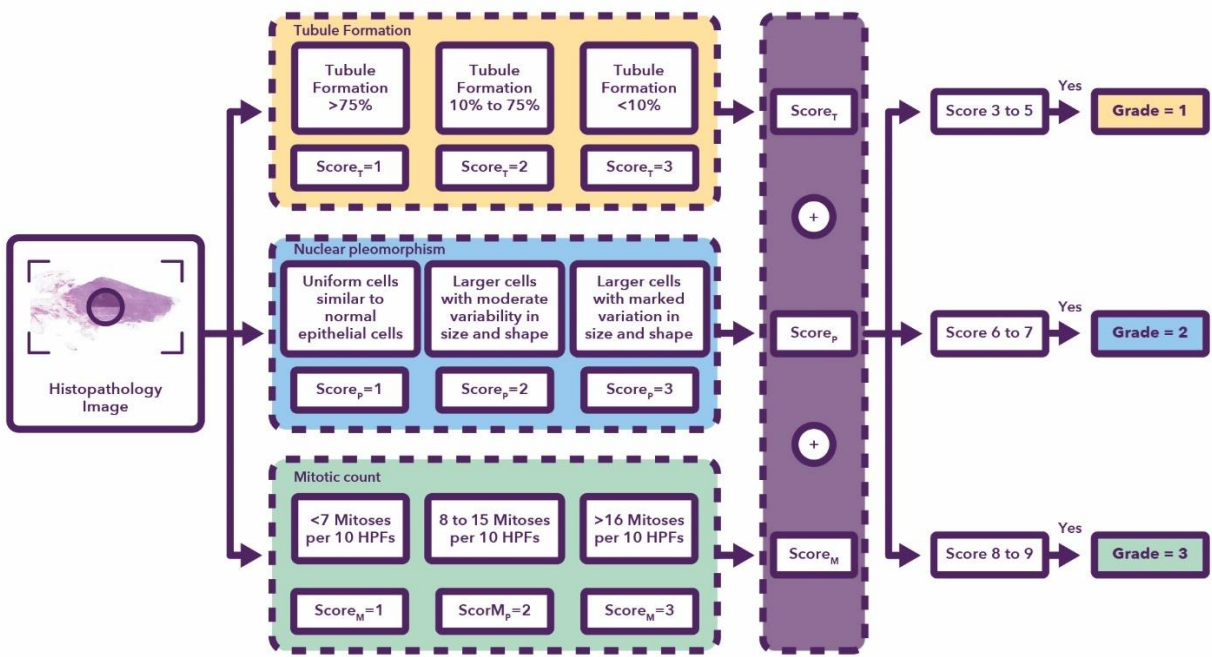

**Fig. 5** A diagnostic algorithm for assigning a grade to a breast cancer histopathology image. (T: tubule formation, P: nuclear pleomorphism, M: mitotic count)

## 3.5 Annotation constructs

Three main annotation constructs: rectangle, point, and polygon, were used to annotate different structures as regions of interest, cell points, and region boundaries. In total 10,731 rectangles, 509,591 points, and 194,717 polygons were used for annotations (including both H&E and IHC). Further detail of the use of constructs is provided in supplement Table S12.

## 3.6 Degree of annotations

Annotations were completed in an exhaustive manner: all the different types of regions and cells defined in the data dictionary within a bounding box were annotated by the pathologists. To quantify this, the proposed exhaustiveness metric was used to identify non-exhaustive annotations. Supplement Fig. S11 shows the comparison of the exhaustiveness versus diversity of H&E region-level annotations for individual and consensus boxes for a subset (n=40) of cases. The diversity here represents how many different types of regions were annotated in a region box. The overall exhaustiveness of the annotations is high with most of the boxes having higher than 80% exhaustiveness for individual boxes (represented by red dots) and above 60% exhaustiveness for consensus boxes (represented by blue dots). Similarly, most of the boxes contain between 3 to 6 types of regions. Supplement Fig. S5 shows examples of exhaustive region and cell annotations.

**Table 2** Examples of number of annotations at different levels. Case-level (grades), Region-level (different regions were annotated using a polygon. Number of region annotation as well as the area in mm$^2$ is listed), Cell-level (different types of cells were annotated using a dot/point).

| | Case-level | | | Region-level | | | | | | Cell-level | | |
|---|---|---|---|---|---|---|---|---|---|---|---|---|
| | H&E | PR | Ki67 | H&E | | PR | | Ki67 | | H&E | PR | Ki67 |
| **Breast** | 1226 | 1143 | 1121 | Count | Area (mm$^2$) | Count | Area | Count | Area | 249834 | 204317 | 57388 |
| | | | | 81262 | 2215.37 | 32809 | 716.80 | 27490 | 864.81 | | | |
| **Colon** | 5000 | NA | | 19848 | 268.36 | NA | | | | 57480 | NA | |

## 3.7 Phases of annotation

To familiarize the multi-disciplinary team with the definitions of different structures and identify any potential annotation issues, two workshops were held where pathologists and ML experts participated. For this purpose, a small set of representative cases (those showing stain variations, tissue size, stain types, and structural diversity) were assigned for a trial annotation to pathologists in groups. This pilot phase not only helped in identifying useful structures but also helped in annotation workload estimation. Once the pathologist team was acquainted with the protocol and the tools, the annotations proceeded in a phased manner, starting with case-level labels. Diagnostic/prognostic records were checked and were compiled into a single structured form as clinical data is often unstructured and maybe distributed across disparate medical information systems. Relevant cells and regions were then annotated to help with detailed supervision of the ML methods.

## 3.8 Annotation software

Due to the distributive nature of the project the locations and time constraints of the pathologists, an open-source online annotation software (Digital Slide Archive (DSA) (39)) was customized and extended and adapted into the WArwick Slide Archive and collaBoration Interface (WASABI). As the current version of the DSA software lacked some features such as workload distribution, an annotation download interface, annotation status (as a leader board), grouping of annotations a frontend gateway was built for this purpose. Annotations were saved in JSON format. To add further features such as ML results visualization, side-by-side image comparison, integration of interactive annotations, and to incorporate other features requested by pathologists and ML experts, an advanced version (WASABI+) is currently under development.

## 3.9 Interactive/active annotations

As annotation of cells/nuclei boundaries is very time consuming, we used NuClick (22), a unified interactive segmentation framework for histopathological images to generate such nuclei segmentation masks. It requires only one point for delineating nuclei and cells and a squiggle for outlining glands or larger areas. Providing a dot inside a nucleus and cell or drawing a squiggle inside a gland is fast, easy, and combined with good performance of the model, makes NuClick user-friendly. Also, it allows for converting existing low-level annotations (points) into dense annotations (segmentation masks). Supplement Fig. S8 illustrates example outputs of NuClick annotation in response to low-level guidance of annotator for nuclei boundary segmentation. Using this framework, we generated a total of 124,624 H&E and 109,862 PR cell boundaries for the BraCe project.

## 3.10 Workload distribution

Based on the pilot phase and some initial annotations, the workload distribution was estimated. The pathologists were put into three teams with a mix of experience in each team, and the slides for annotation were split equally amongst the teams. As the project progressed, some annotators completed their work more quickly than other staff of different time pressures and clinical commitments. Groups of two pathologists with a mix of experiences were formed so that consensus annotations could be done.

## 3.11 Quality review

Throughout the steps of our proposed annotation workflow, a regular quality review for each stage was done. For example, any change/removal/addition of a structure was documented in the data dictionary for future reference. All the annotation analysis reports were recorded, and an online log of annotation issues was maintained.

### 3.11.1 Image Quality analysis

The results of ImageQC pipeline on WSIs with pen-marking, coverslip edges and blurriness are shown in supplement Fig. S12. It can be observed that the blue, green, and dark green pen-markings, shadow region on the coverslip edges (Fig. S12 A and B) have been adequately removed during the workflow. Similarly, Figs. S12 C shows the detection of pen-marking and blurriness in two other WSIs.

### 3.11.2 Annotation Quality Analysis

The bar chart in supplement Fig. S13 shows the annotations of different types of regions in terms of counts and area (in $mm^2$). This type of analysis helps in prioritizing the regions for annotations. For example, if a region such as 'usual ductal hyperplasia' is of high importance but there are relatively few annotations of this region type then it can be elevated in the annotation priority list. Supplement Fig. S14 shows inter-annotator agreement for sample regions in terms of Jaccard Similarity Index versus the area of the regions. It can be observed that though some regions such as fibrocystic change and lobular neoplasia are relatively low in terms of annotated area (0.674 $mm^2$ and 0.539 $mm^2$, respectively), agreement on them is still very good (> 80%).

### 3.11.3 Pathologist agreement

Fig. 6-A shows examples of variability among pathologists for two region types of tumor and tumor-associated stroma. Inter-pathologist variability might be a result of annotator's bias, experience, judgment, ambiguous definitions in the data dictionary, or the difficulty in delineating some regions.

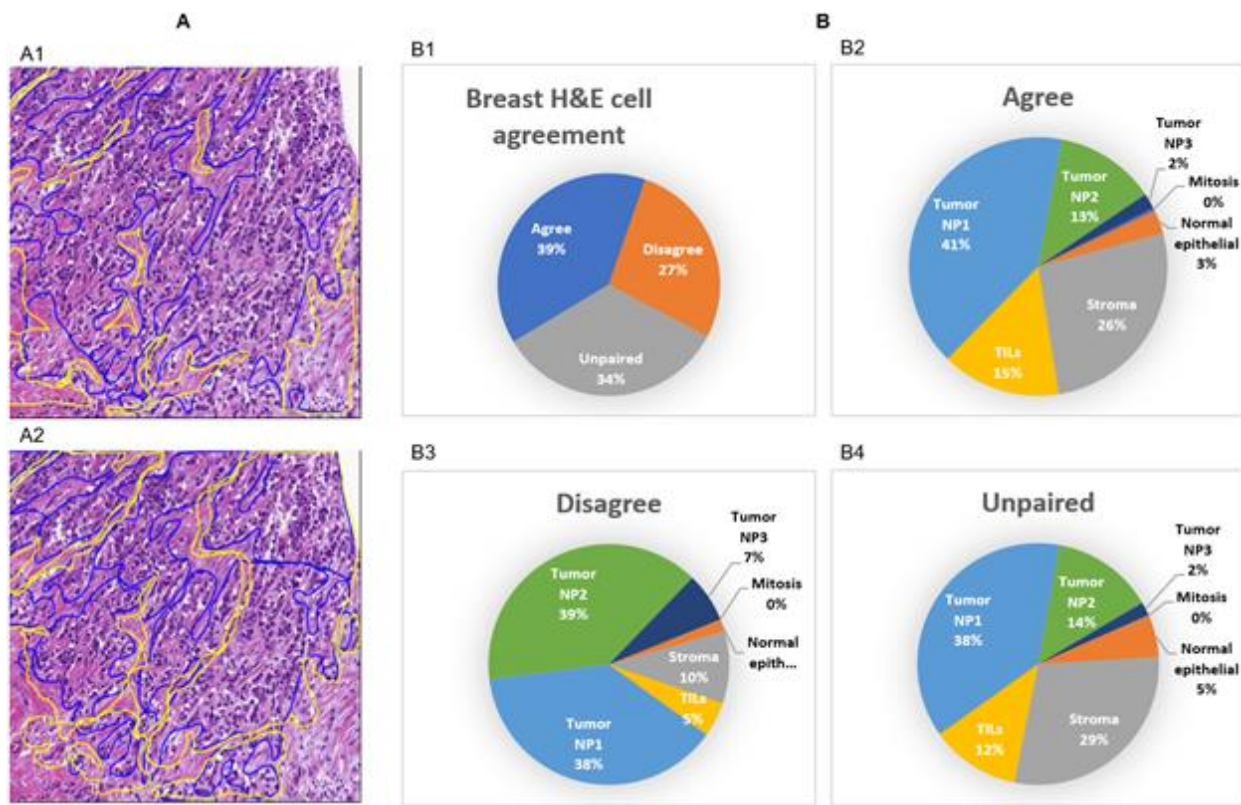

**Fig. 6** A. An example of annotation variability between two pathologists (A1 Tumor-associated stroma, A2 Tumor). Annotations by Pathologist 1 (blue), Pathologist 2 (yellow). B. Mean percentage of cells (B1) on which two pathologists agreed (B2), disagreed (B3) and missed by one pathologist (B4) in breast H&E cell annotations. NP (nuclear pleomorphism)

Inter-pathologist agreement/disagreement on cell-level annotations is shown in Fig. 6-B. To measure the agreement, point annotations within a radius of 12 pixels at 40× magnification (approx. 0.25 μm per pixel) were considered as annotation for the same cell. It can be observed that it was quite common for the pathologists to miss some cells (34% Fig. 6-B1), even when exhaustively annotating inside a bounding box. Specific types of cell nuclei annotated by one pathologist, but missed by another, include Tumor NP1 (nuclear pleomorphism 1) and NP2 and stroma cells (Fig. 6-B4). The highest disagreement is exhibited for tumor cells pleomorphism (Fig. 6-B3). It is important to discuss such issues of disagreement in pathologists' review meetings to reach a consensus (for example, discussing and updating the features of pleomorphism).

To further analyze the inter-pathologist discordance on different cell types supplement Table S13 presents the confusion matrix. It is evident that the different categories of NP are quite challenging to identify, especially differentiating NP1 from NP2 (4,699 disagreed).

## 3.12 Annotation usage

As a demonstrative example, we present our results on using the annotations gathered in our BraCe project for development of a ML model for classification of different breast cells in H&E WSIs. For this purpose our in-house HoVerNet cell classification and segmentation model (40), which was pre-trained on the PanNuke dataset (41), was fine-tuned to classify breast H&E cells by using a subset of the annotation in a 70-30 holdout training/validation protocol. The number of cell annotations included 20,822 tumor, 3,134 TILs, 7,119 stromal, and 1 528 normal epithelial cells. A marked improvement in classification results, from a macro F1 score of 0.53 to 0.79 (supplement Table S14), with the use of the annotations clearly demonstrating the usefulness of annotations collected in this manner. The cell-level annotations were also used in an interactive manner via NuClick to generate segmentation masks for different types of cells. We generated a total of 124,624 and 109,862 nuclear masks for breast H&E and PR images, respectively. The annotation analysis also helped in prioritizing annotation of different regions and cells.

# 4. Discussion and Conclusions

Although there has been some work on standardizing the ontologies of medical terms such as SNOMED CT (42), National Cancer Institute Thesaurus (NCIt) (43), and foundational model of anatomy (FMA) (44), protocols for histopathology images need standardization. To increase the usage of annotated dataset, it is important that a standard annotation protocol agreed between pathologists and the ML team is followed. In the absence of a standard protocol, it can become very difficult to make use of existing annotated datasets. Similarly, interoperability of annotations is a big issue currently. For example, because of the existence of many different image formats and compression models for storing WSIs, an annotation software might not support all the available file formats for different scanners. Furthermore, each annotation software has its own unique way of storing the annotations, XML, JSON, CSV, then a conversion process will be required. To make use of an annotated dataset, one should ideally only require the annotation file and its corresponding data dictionary. There is also a need to standardize the process of annotation reviews by the pathologists so that discordant annotations can be resolved in a systematic manner. In this work, we have developed a first annotation protocol for CPath projects. To make the best use of time-consuming and labor-intensive annotations, and to minimize diagnostic drift and inconsistencies in the nomenclature of annotation structures, it is important to follow standard guideline. To standardize the process for annotating medical images of CPath projects this manuscript presents comprehensive guidelines with relevant examples. To minimize the discrepancies between pathologists in annotations and to make the downstream analysis reproducible, different quality control metrics are proposed to quantitatively assess the annotations and are further discussed in the context of the annotation workflow. Recommendations are made for important regions and standardization of annotations with directions for future.

It is important to note that the annotation strategy should be determined by the task and this would mainly be determined by discussion between the ML and pathology teams. Other approaches of annotations might be considered if conventional cell by cell or region by region annotations are not sufficient for making an assessment (45). In case of ambiguous structures, where the pathologist is not sure about the category, it is advisable to keep a category of 'unknown' regions and cells to avoid noisy annotations for ML model training and further assessment.

In a future study, it would be useful to apply the proposed guidelines to the complete lifecycle of a CPath project to see the effects of some aspects, such as interactive and active annotations for expediting the annotation process, interoperability, and use of non-exhaustive annotations. Similarly, the extension of the data dictionary and the associated annotation schema to other projects is required to see what overhead may be incurred in adaptation of the proposed annotation protocols. To demonstrate the importance of annotations done by domain experts, a comprehensive comparison of strongly supervised and weakly supervised or self-supervised methods is needed. It will also be interesting to see the usefulness of the collected annotations in exploratory algorithms other than diagnostic and prognostic. Furthermore, crowdsourcing might be considered as an alternative annotation strategy. A study on a wide range of pathological structures would be beneficial to compare the quality of annotations from a large pool of non-experts compared to pathologists. Again, a small number of pathologists might be added to benefit a crowdsourcing based annotation learning and refinement process which may help in identifying difficult-to-annotate structures.

## Acknowledgments

Manuel Salto-Tellez and Jacqueline A James are Principal Investigators in PathLAKE at Queens's University Belfast and Clare Verrill is a Principal Investigator in PathLAKE at the University of Oxford — all were involved in generating the PathLAKE programme, including funding.

## Conflict of Interest

The authors have declared no conflicts of interest.

**Ethics Approval and Consent to Participate:**

This study was approved by the Yorkshire & The Humber - Leeds East Research Ethics Committee (REC Reference: 19/YH/0293) under the IRAS Project ID: 266925. Data collected were fully anonymized.

## Author Contributions

Conceptualization: NR and FM, experiments: NW, annotations: IM, MT, AL, AI, HOE, MP, and AK, Data Dictionary: AB, SR, MT, DS, and ER, ImageQC: WL, MB, and SG, interactive annotation section: MJ, software development: YP, and GH, writing (editing): NR, FM, NW, IM, KD, and HS, writing (reviewing): AB, MT, TS, EH, HE, YT, and KG. All authors read and approved the final paper.

## Funding

This paper is supported by the PathLAKE Centre of Excellence for digital pathology and artificial intelligence which is funded from the Data to Early Diagnosis and Precision Medicine strand of the HM Government's Industrial Strategy Challenge Fund, managed and delivered by Innovate UK on behalf of UK Research and Innovation (UKRI). Views expressed are those of the authors and not necessarily those of the PathLAKE Consortium members, the NHS, Innovate UK or UKRI. Grant ref: File Ref 104689/application number 18181.

## Data Availability Statement

All annotations and the corresponding annotation protocols will be made available upon completion of PathLAKE project.

# References

1.  Srinidhi CL, Ciga O, Martel AL. Deep neural network models for computational histopathology: A survey. Med Image Anal. 2021;67:101813.

2.  Hayakawa T, Prasath VBS, Kawanaka H, Aronow BJ, Tsuruoka S. Computational Nuclei Segmentation Methods in Digital Pathology: A Survey. Arch Comput Methods Eng. 2021;28(1):1–13.

3.  BIGPICTURE. [Internet], [cited 8 June 2021]. Available from: https://www.imi.europa.eu/projects-results/project-factsheets/bigpicture

4.  Patel T. Review of "Digital Pathology" by Liron Pantanowitz and Anil V Parwani. J Pathol Inform. 2017 Jan 1;8(1):37.

5.  Hamilton PW, Bankhead P, Wang Y, Hutchinson R, Kieran D, McArt DG, et al. Digital pathology and image analysis in tissue biomarker research. Methods. 2014;70(1):59–73.

6.  Robertson S, Azizpour H, Smith K, Hartman J. Digital image analysis in breast pathology-from image processing techniques to  artificial intelligence. Transl Res. 2018 Apr;194:19–35.

7.  Bokhorst J-M, Pinckaers H, van Zwam P, Nagtegaal I, van der Laak J, Ciompi F. Learning from sparsely annotated data for semantic segmentation in histopathology images. Proceedings of The 2nd International Conference on Medical Imaging with Deep Learning. London, United Kingdom: PMLR; 2019. p. 84–91. (Proceedings of Machine Learning Research; vol. 102).

8.  Yamaguchi T, Mukai H, Akiyama F, Arihiro K, Masuda S, Kurosumi M, et al. Inter-observer agreement among pathologists in grading the pathological response to neoadjuvant chemotherapy in breast cancer. Breast Cancer. 2018;25(1):118–25.

9.  PinnacleCare. The human cost and financial impact of misdiagnosis, 2019 [Internet] [cited 8 June 2021]. Available from: https://www.pslhub.org/learn/patient-safety-in-health-and-care/diagnosis/diagnostic-error/white-paper-the-human-cost-and-financial-impact-of-misdiagnosis-2016-r2551/

10. Yu K-H, Berry GJ, Rubin DL, Ré C, Altman RB, Snyder M. Association of Omics Features with Histopathology Patterns in Lung Adenocarcinoma. Cell Syst. 2017;5(6):620-627.e3.

11. He B, Bergenstråhle L, Stenbeck L, Abid A, Andersson A, Borg Å, et al. Integrating spatial gene expression and breast tumour morphology via deep learning. Nat Biomed Eng. 2020;4(8):827–34.

12. Park S, Parwani A V, Aller RD, Banach L, Becich MJ, Borkenfeld S, et al. The history of pathology informatics: A global perspective. J Pathol Inform. 2013 May 30;4:7.

13. Robboy SJ, Weintraub S, Horvath AE, Jensen BW, Alexander CB, Fody EP, et al. Pathologist workforce in the United States: I. Development of a predictive model to examine factors influencing supply. Arch Pathol Lab Med. 2013 Dec;137(12):1723–32.

14.     Wahab N, Khan A. Multifaceted fused-CNN based scoring of breast cancer whole-slide histopathology images. Appl Soft Comput. 2020;97:106808.

15.     Amgad M, Atteya LA, Hussein H, Mohammed KH, Hafiz E, Elsebaie MAT, et al. NuCLS: A scalable crowdsourcing, deep learning approach and dataset for nucleus classification, localization and segmentation. Preprint at https://arxiv.org/ftp/arxiv/papers/2102/2102.09099.pdf (2021).

16.     Awan R, Sirinukunwattana K, Epstein D, Jefferyes S, Qidwai U, Aftab Z, et al. Glandular Morphometrics for Objective Grading of Colorectal Adenocarcinoma Histology Images. Sci Rep. 2017;7(1):16852.

17.     Esteva A, Kuprel B, Novoa RA, Ko J, Swetter SM, Blau HM, et al. Dermatologist-level classification of skin cancer with deep neural networks. Nature. 2017;542(7639):115–8.

18.     Jing L, Tian Y. Self-supervised Visual Feature Learning with Deep Neural Networks: A Survey. IEEE Trans Pattern Anal Mach Intell. 2020;1.

19.     Campanella G, Hanna MG, Geneslaw L, Miraflor A, Werneck Krauss Silva V, Busam KJ, et al. Clinical-grade computational pathology using weakly supervised deep learning on whole slide images. Nat Med. 2019;25(8):1301–9.

20.     Zhou Z-H. A brief introduction to weakly supervised learning. Natl Sci Rev. 2017;5(1):44–53.

21.     Qu H, Wu P, Huang Q, Yi J, Riedlinger GM, De S, et al. Weakly Supervised Deep Nuclei Segmentation using Points Annotation in Histopathology Images. Proceedings of The 2nd International Conference on Medical Imaging with Deep Learning [Internet]. London, United Kingdom: PMLR; 2019. p. 390–400. (Proceedings of Machine Learning Research; vol. 102).

22.     Alemi Koohbanani N, Jahanifar M, Zamani Tajadin N, Rajpoot N. NuClick: A deep learning framework for interactive segmentation of microscopic images. Med Image Anal. 2020;65:101771.

23.     Hou L, Agarwal A, Samaras D, Kurc TM, Gupta RR, Saltz JH. Robust Histopathology Image Analysis: To Label or to Synthesize? In: Proceedings of the IEEE/CVF Conference on Computer Vision and Pattern Recognition (CVPR). 2019.

24.     Baselli G, Codari M, Sardanelli F. Opening the black box of machine learning in radiology: can the proximity of annotated cases be a way? Eur Radiol Exp. 2020 May 5;4(1):30.

25.     Grünberg K, Jimenez-del-Toro O, Jakab A, Langs G, Salas Fernandez T, Winterstein M, et al. Annotating Medical Image Data. Cloud-Based Benchmarking of Medical Image Analysis. Cham: Springer International Publishing; 2017. p. 45–67.

26.     Willemink MJ, Koszek WA, Hardell C, Wu J, Fleischmann D, Harvey H, et al. Preparing Medical Imaging Data for Machine Learning. Radiology. 2020;295(1):4–15.

27.     Jones SR, Carley S, Harrison M. An introduction to power and sample size estimation.

Emerg Med J. 2003;20(5):453–8.

28. Figueroa RL, Zeng-Treitler Q, Kandula S, Ngo LH. Predicting sample size required for classification performance. BMC Med Inform Decis Mak. 2012 Feb 15;12:8.

29. Kanavati F, Toyokawa G, Momosaki S, Rambeau M, Kozuma Y, Shoji F, et al. Weakly-supervised learning for lung carcinoma classification using deep learning. Sci Rep. 2020;10(1):9297.

30. Quellec G, Cazuguel G, Cochener B, Lamard M. Multiple-Instance Learning for Medical Image and Video Analysis. IEEE Rev Biomed Eng. 2017;10:213–34.

31. Babenko B. Multiple instance learning: algorithms and applications. View Artic PubMed/NCBI Google Sch. 2008;1–19.

32. Li S, Liu Y, Sui X, Chen C, Tjio G, Ting DSW, et al. Multi-Instance Multi-Scale CNN for Medical Image Classification BT - Medical Image Computing and Computer Assisted Intervention – MICCAI 2019. In: Shen D, Liu T, Peters TM, Staib LH, Essert C, Zhou S, et al., editors. Cham: Springer International Publishing; 2019. p. 531–9.

33. Wieneke AE, Bowles EJA, Cronkite D, Wernli KJ, Gao H, Carrell D, et al. Validation of natural language processing to extract breast cancer pathology procedures and results. J Pathol Inform. 2015 Jun 23;6:38.

34. Kather JN, Heij LR, Grabsch HI, Loeffler C, Echle A, Muti HS, et al. Pan-cancer image-based detection of clinically actionable genetic alterations. Nat Cancer. 2020;1(8):789–99.

35. Esteva A, Wal D van der, Jhun I, Laklouk I, Nirschl J, Richer L, et al. Biological data annotation via a human-augmenting AI-based labeling interface. Research Square; 2021.

36. Cheung CC, Torlakovic EE, Chow H, Snover DC, Asa SL. Modeling complexity in pathologist workload measurement: the Automatable Activity-Based Approach to Complexity Unit Scoring (AABACUS). Mod Pathol. 2015;28(3):324–39.

37. Janowczyk A, Zuo R, Gilmore H, Feldman M, Madabhushi A. HistoQC: An Open-Source Quality Control Tool for Digital Pathology Slides. JCO Clin cancer informatics. 2019 Apr;3:1–7.

38. Crete F, Dolmiere T, Ladret P, Nicolas M. The blur effect: perception and estimation with a new no-reference perceptual blur metric. In: Human vision and electronic imaging XII. 2007. p. 64920I.

39. Gutman DA, Khalilia M, Lee S, Nalisnik M, Mullen Z, Beezley J, et al. The Digital Slide Archive: A Software Platform for Management, Integration, and Analysis of Histology for Cancer Research. Cancer Res. 2017 Nov 1;77(21):e75 LP-e78.

40. Graham S, Vu QD, Raza SEA, Azam A, Tsang YW, Kwak JT, et al. Hover-Net: Simultaneous segmentation and classification of nuclei in multi-tissue histology images. Med Image Ana. 2019;58:101563.

41.     Gamper J, Alemi Koohbanani N, Benet K, Khuram A, Rajpoot N. PanNuke: An Open Pan-Cancer Histology Dataset for Nuclei Instance Segmentation and Classification. In: Reyes-Aldasoro CC, Janowczyk A, Veta M, Bankhead P, Sirinukunwattana K, editors. Cham: Springer International Publishing; 2019. p. 11–9.

42.     Donnelly K. SNOMED-CT: The advanced terminology and coding system for eHealth. Stud Health Technol Inform. 2006;121:279–90.

43.     NCI Thesaurus (NCIt). [Internet], [cited 8 June 2021]. Available from: https://ncithesaurus.nci.nih.gov/

44.     Musen MA, Noy NF, Shah NH, Whetzel PL, Chute CG, Story M-A, et al. The National Center for Biomedical Ontology. J Am Med Inform Assoc. 2012;19(2):190–5.

45.     Chatrian A, Colling RT, Browning L, Alham NK, Sirinukunwattana K, Malacrino S, et al. Artificial intelligence for advance requesting of immunohistochemistry in diagnostically uncertain prostate biopsies. Mod Pathol. 2021;

# Supplementary material

**"Semantic annotation for computational pathology: Multidisciplinary experience and best practice recommendations"**

Noorul Wahab et al. June, 2021

## S1. Materials and methods

## S1.1 Annotation data dictionary

**Table S1** Description of bounding boxes used for region annotations in PathLAKE annotation data dictionary.

| Term | Definition |
|---|---|
| Cell Box Individual (Box_Cell_20x) | A box in which each cell is annotated by one annotator only. A box drawn at 20x magnification to cover a visual field. |
| Region Box Individual (Box_Region_5x) | A box in which each region is annotated by one annotator only. A box drawn at 5x magnification to cover a visual field. |
| Cell Box Consensus (Box_All_Cell_20x) | A box in which each cell is annotated by multiple annotators. A box drawn at 20x magnification to cover a visual field. |
| Region Box Consensus (Box_All_Region_5x) | A box in which each region is annotated by multiple annotators. A box drawn at 5x magnification to cover a visual field. |
| Cell Box Individual Special (Box_Special_20x) | A box in which each cell of special type (e.g., mitosis) is annotated by one annotator only. A box of an appropriate size drawn at 20x magnification. Every cell of the special type inside this box is annotated by a single annotator only. It is used for mitosis, normal epithelial, and tumor cells. |
| Region Box Individual Special (Box_Special_5x) | A box in which each region of special type (e.g., DCIS) is annotated by one annotator only. A box of an appropriate size drawn at 5x magnification. Every region of the special type inside this box is annotated by a single annotator. It is used for DCIS, vascular channels, lymphoid follicles, normal TDLUs, etc. |
| Cell Box Consensus Special (Box_ALL_Special_20x) | A box in which each cell of special type (e.g., mitosis) is annotated by multiple annotators. A box of an appropriate size drawn at 20x magnification. Every cell of the special type inside this box is annotated by multiple |

| | annotators.  It is used for mitosis, normal epithelial, and tumor cells. |
|---|---|
| Region Box Consensus Special (Box_ALL_Special_5x) | A box in which each and every region of special type (e.g., DCIS) is annotated by multiple annotators. A box of an appropriate size drawn at 5x magnification. Every region of the special type inside this box is annotated by multiple annotators.  It is used for DCIS, vascular channels, lymphoid follicles, normal TDLUs, etc. |

Recommendations on semantic annotation for computational pathology

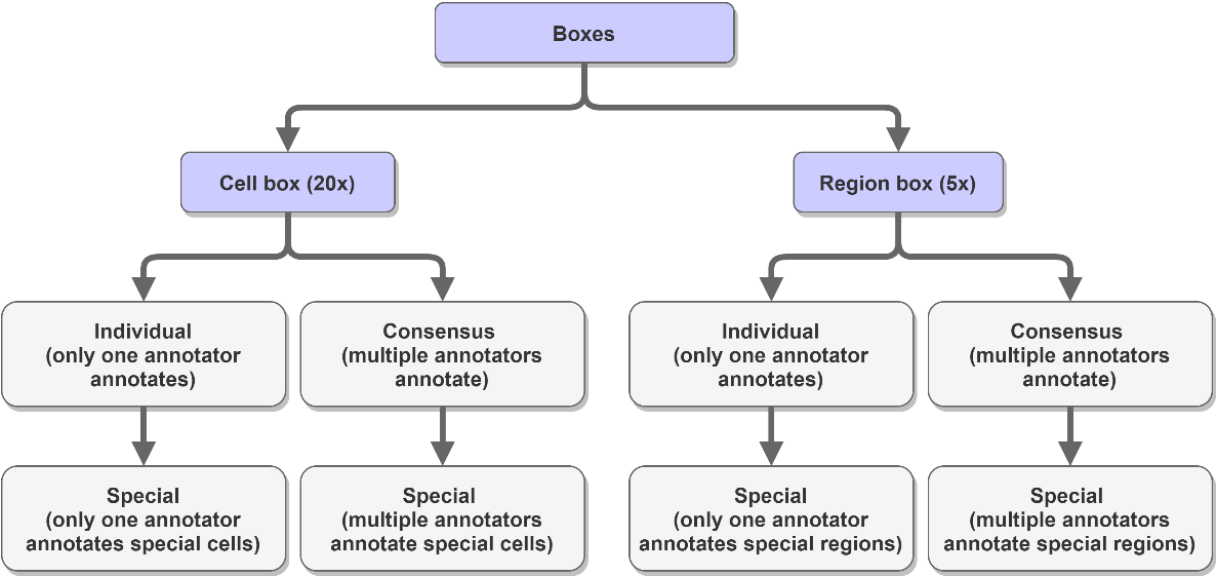

**Fig. S1** Sample tree diagram of different bounding boxes from PathLAKE annotation data dictionary

**Table S2** Description of region annotations for breast H&E from PathLAKE annotation data dictionary

| Region name | Description | Example |
|---|---|---|
| **Normal** | | |
| Normal TDLUs | Normal terminal ductal lobular units (TDLUs) | 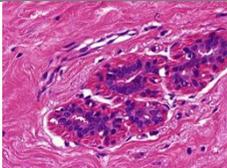 |
| Normal skeletal muscles | Normal skeletal muscles | 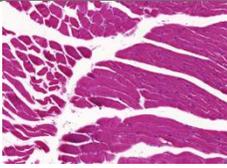 |
| Normal nerves | Normal Nerve bundles | 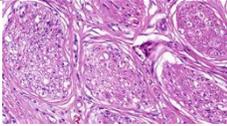 |
| Normal vascular channels | Normal vascular channels (blood vessels or lymphatic channels) | 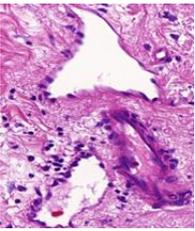 |
| **Benign/non-malignant lesions** | | |
| Benign papilloma | Benign papilloma are benign tumors that grow from epithelial tissue and form outward finger-like fronds with fibrovascular cores. | 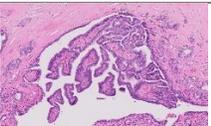 |
| Fibroadenoma | Fibroadenomas are made up of overgrowth of both glandular tissue and stroma tissue. | 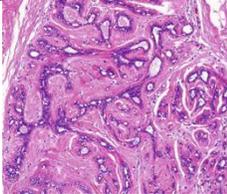 |
| Fibrocystic change | Fibrocystic change. Dilated cysts along with areas of fibrosis. | 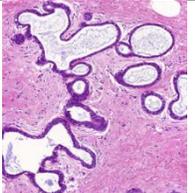 |
| Usual ductal hyperplasia | Hyperplasia is an overgrowth of the cells that line the breast ducts. Cells are not neoplastic. | 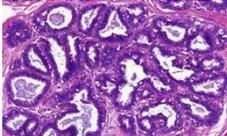 |
| Benign others | Any other benign lesion that can be present in a given slide such as radial scar, microglandular adenosis, lactational changes, etc. | 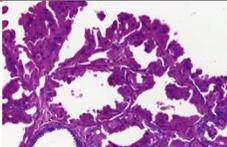 |

**DCIS/LCIS**

| | | |
|---|---|---|
| DCIS | Ductal carcinoma in situ (DCIS) is a non-invasive form of BC which is defined as neoplastic epithelial cells proliferating within the mammary ducts, which have not breached the basement membrane nor invaded surrounding tissues. | 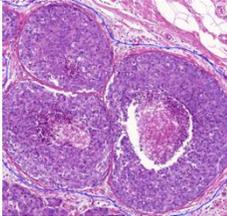 |
| Lobular neoplasia (LN) | Lobular neoplasia is a condition in which atypical cells are found in the terminal lobules of the breast. This category includes atypical lobular hyperplasia and lobular carcinoma in situ (LCIS). | 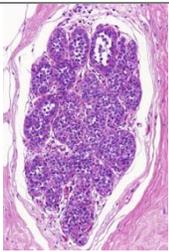 |

**Stroma**

| | | |
|---|---|---|
| Stroma fibroblastic | Fibroblasts are connective tissue which produces collagen and other fibers. This type of stroma is found around non-malignant lesions. | 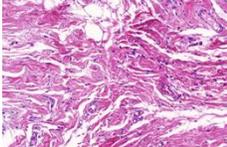 |
| Stroma tumor associated | Complex Stroma (tumor associated) mainly consists of the basement membrane, fibroblasts, myofibroblasts extracellular matrix, immune cells, and vasculature. This type of stroma is associated with malignant lesions. | 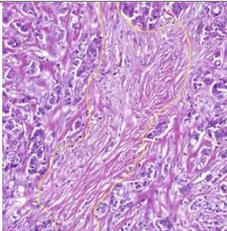 |

| Stroma fibrofatty | Fibrofatty stroma | 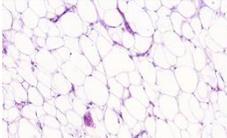 |
| --- | --- | --- |

| **Tumor** | | |
| --- | --- | --- |
| Tumor tubules acini | Tubule/Acinar formation. The tumor exhibits tubule formation like normal breast. | 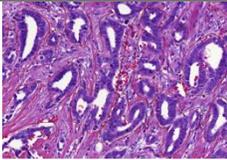 |
| Tumor non tubular | Tumor (Non-tubular) Sheets/cords, nests, single cells | 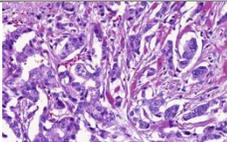 |
| Tumor mucinous | It is a rare form of invasive ductal carcinoma made up of malignant epithelial cells that float in pools of mucin. | 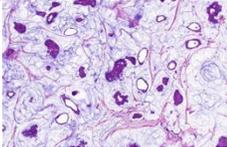 |
| Tumor invasive mixed | Invasive tumor mixed pattern. This includes more than one invasive tumor pattern in a single lesion, for example ductal and lobular in a single image. | 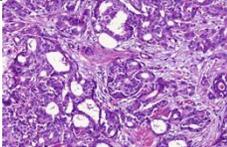 |

| **Other** | | |
| --- | --- | --- |
| Lymphoid follicle | Aggregates of lymphocytic and histiocytic infiltrate plasma cells. Often shows germinal cells formation. | 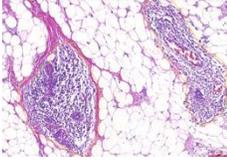 |
| Lympho-vascular invasion | It is malignant cells within blood vessels and/or lymphatics. | 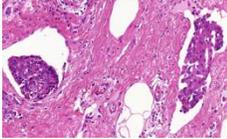 |
| Perineural invasion | Perineural invasion (PNI) is described when cancer cells encroach along nerves. | 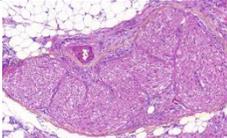 |

| Exclusion | Areas of exclusion include folding, Tissue Micro Array site, any area where no tissue is observed, ink, out of focus regions and air bubbles) | 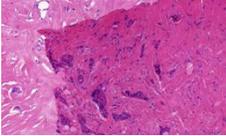 |
| Extra features | Extra-features such as biopsy site changes, calcifications, necrosis, and hemorrhage | 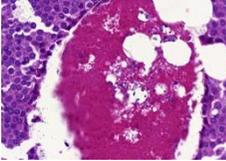 |

**Table S3** Description of cell annotations for breast H&E from PathLAKE annotation data dictionary

| Cell name | Description | Example |
|---|---|---|
| **Mitosis** | Mitosis either typical (prophase, metaphase, anaphase, and telophase) or atypical (anaphase bridge, multipoles, and asymmetrical shape) | 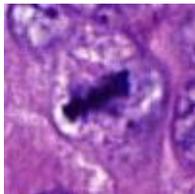 |
| **Normal epithelial/ myoepithelial** | Normal epithelial cells are the innermost layer of bilayered ductolobular system and<br><br>Myoepithelium: outermost layer resting on a basement membrane | 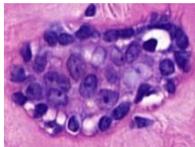 |
| **Epithelial NP1 (nuclear pleomorphism 1)** | Tumor cells which are similar in size to normal epithelial cells. | 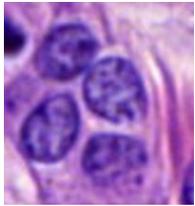 |
| **Epithelial NP2 (nuclear pleomorphism 2)** | Tumor cells with moderate variability in size and shape. They are larger in size with open vesicular nuclei and visible nucleoli. | 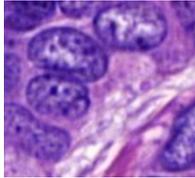 |
| **Epithelial NP3 (nuclear pleomorphism 3)** | Tumor cells with high variability in size and shape. | 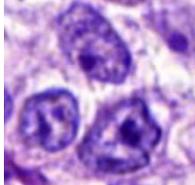 |

**Table S4** Annotation styles for regions (Breast H&E) which could be used to generate a style file (such as in JSON format) to be used with an annotation software. (From PathLAKE annotation data dictionary)

| Type | Style name | Annotation shape | Line Color | RGB/HEX Values |
|---|---|---|---|---|
| Normal TDLUs | HE_Region_normal_TDLUs | Polygon | Green | 0,128,0 #008000 |
| Normal skeletal muscles | HE_Region_normal_skeletal_ms | Polygon | Light Green | 144,238,144 #90EE90 |
| Normal nerves | HE_Region_normal_nerves | Polygon | Dark Sea Green | 143,188,143 #8FBC8F |
| Normal vascular channels | HE_Region_normal_vascular_ch | Polygon | Lavender | 230,230,250 #E6E6FA |
| Papillomas | HE_Region_benign_papillomas | Polygon | Deep Pink | 255,20,147 #FF1493 |
| Fibroadenoma | HE_Region_benign_fibroadenoma | Polygon | Aqua | 0,255,255 #00FFFF |
| Fibrocystic change | HE_Region_benign_fibrocystic_change | Polygon | Lime | 0,255,0 #00FF00 |
| Usual ductal hyperplasia | HE_Region_benign_hyperplasia | Polygon | Coral | 255,127,80 #FF7F50 |
| Benign others | HE_Region_benign_others | Polygon | Cadet Blue | 95,158,160 #5F9EA0 |
| DCIS | HE_Region_DCIS | Polygon | Blue | 0,0,255 #0000FF |
| Lobular neoplasia | HE_Region_lobular_neoplasia | Polygon | Light Sky Blue | 135,206,250 #87CEFA |
| Stroma fibroblastic | HE_Region_stroma_fibroblastic | Polygon | Light Yellow | 255,255,102 #FFFF66 |
| Stroma tumor associated | HE_Region_stroma_tumor_associated | Polygon | Yellow | 255,255,0 #FFFF00 |
| Stroma fibrofatty | HE_Region_stroma_fibrofatty | Polygon | Crimson | 220,20,60 #DC143C |
| Tumor mucinous | HE_Region_tumor_mucinous | Polygon | Orange Red | 255,69,0 #FF4500 |
| Tumor invasive mixed | HE_Region_ tumor_invasive_mixed | Polygon | Light Red | 224,102,102 #E06666 |
| Tumor tubules acini | HE_Region_tumor_tubules_acini | Polygon | Dark Red | 139,0,0 #8B0000 |
| Tumor non tubular | HE_Region_tumor_non_tubular | Polygon | Red | 255,0,0 #FF0000 |
| Lymphoid | HE_Region_lymphoid | Polygon | Orange | 255,165,0 #FFA500 |
| Lympho-vascular invasion | HE_Region_LVI | Polygon | Dark Magenta | 139,0,139 #8B008B |
| Perineural invasion | HE_Region_PNI | Polygon | Gold | 255,215, 0 #FFD700 |
| Exclusion | HE_Region_exclusion | Polygon | | 0,0,0 #000000 |
| Extra features | HE_Region_extra_features | Polygon | Cream White | 255, 253, 208 #FFFDD0 |

trueRecommendations on semantic annotation for computational pathology

## S1.2 Annotation software

**Table S5** Comparison of different annotation software

| Software | AIDA (49) 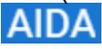 | ASAP (49) 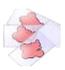 | Cytomine (50) 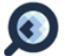 | DSA (42) 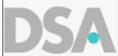 | Omero (51) 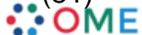 | QuPath (52) 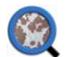 |
|---|---|---|---|---|---|---|
| Supported Annotation Constructs | Line, dot, rectangle, polygon, path, paint | Line, dot, rectangle, polygon, spline | Dot, rectangle, polygon, line, circle | Line, dot, rectangle, polygon | Line, dot, rectangle, polygon, text | Line, dot, rectangle, polygon |
| Internet accessibility | ✓ | ✗ | ✓ | ✓ | ✓ | ✗ |
| Supported Image formats | dzi | OpenSlide supported format (ndpi, svs, tif, mrxs, vms etc.) | mrxs, tif, tiff, vms, vsi, ndpi, svs, bif | mrxs, tiff, ndpi, svs | Bio-Formats (svs, h5, ndpi, tif, jp2, dicom etc.) | OpenSlide and Bio-Formats |
| Annotation format | JSON | XML | JSON | JSON | JSON | JSON |
| Open source | ✓ | ✓ | ✓ | ✓ | ✓ | ✓ |
| ML integration | ✗ | ✗ | ✓ | ✓ | Via third-party software | ✓ |
| Overlay visualization | ✗ | ✓ | ✓ | ✗ | ✓ | ✓ |
| Side-by-side viewing | ✗ | ✗ | ✓ | ✗ | ✓ | ✗ |

## S1.3 Region and cell annotations

**Table S6** Region-level annotations for breast cancer H&E WSIs

| Region | Count | Area (mm$^2$) |
|---|---|---|
| Fibroadenoma | 2 | 1.19 |
| Fibrocystic change | 34 | 17.32 |
| Usual ductal hyperplasia | 100 | 9.84 |
| Benign others | 36 | 11.34 |
| Papillomas | 8 | 2.76 |
| DCIS | 2714 | 395.21 |
| Exclusion | 1,229 | 43.26 |
| Extra features | 638 | 43.05 |
| Lobular neoplasia | 421 | 16.92 |
| Lympho-vascular invasion (LVI) | 303 | 20.82 |
| Lymphoid | 443 | 36.54 |
| Normal nerves | 275 | 9.13 |
| Normal skeletal muscle | 212 | 68.60 |
| Normal TDLUs | 4,383 | 142.77 |
| Normal vascular channels | 7,058 | 30.38 |
| Perineural invasion | 20 | 1.22 |
| Stroma fibroblastic | 1,628 | 151.51 |
| Stroma fibrofatty | 4,851 | 247.54 |
| Stroma tumor associated | 11,050 | 209.11 |
| Tumor invasive mixed | 124 | 9.52 |
| Tumor mucinous | 305 | 360.66 |
| Tumor non tubular | 42,750 | 359.52 |
| Tumor tubules acini | 5,293 | 27.04 |
| Total | **79,494** | **2,215.37** |

**Table S7** Region-level annotations for breast cancer PR WSIs

| Region | Count | Area (mm$^2$) |
|---|---|---|
| DCIS | 1,740 | 290.56 |
| Exclusion | 589 | 43.68 |
| Lymphoid Follicles | 381 | 50.56 |
| Normal | 3,687 | 113.80 |
| Tumor | 26,407 | 218.16 |
| **Total** | **32,804** | **716.76** |

**Table S8** Region-level annotations for breast cancer Ki67 WSIs

| Region | Count | Area (mm$^2$) |
|---|---|---|
| DCIS | 1 586 | 330.26 |
| Exclusion | 589 | 59.20 |
| Ki67 hotspot* | 74 | 18.26 |
| Lymphoid Follicles | 388 | 58.90 |
| Normal | 3343 | 105.48 |
| Tumor | 21,510 | 292.72 |
| **Total** | **27,490** | **864.81** |

*Not annotated in all images. Only few examples annotated to compare between the performance of ML algorithm and pathologists. However, heatmaps created by the ML model will highlight these hotspots easier and more accurate that human eyeballing.

**Table S9** Cell-level annotations for breast cancer H&E WSIs

| Cell type | Count |
|---|---|
| **Mitosis** | 5,354 |
| **Normal epithelial** | 43,852 |
| **Stroma** | 47,123 |
| **Tumor infiltrating lymphocytes (TILs)** | 29,581 |
| **Tumor Nuclear pleomorphism 1 (NP1)** | 72,492 |
| **Tumor NP2** | 40,590 |
| **Tumor NP3** | 10,842 |
| **Total** | **249,834** |

**Table S10** Cell-level annotations for breast cancer PR WSIs

| Tumor cell nuclear staining | Count |
|---|---|
| Negative | 56,210 |
| Positive moderate intensity | 57,140 |
| Positive strong intensity | 46,133 |
| Positive weak intensity | 44,834 |
| **Total** | **204,317** |

**Table S11** Cell-level annotations for breast cancer Ki67 WSIs

| Cell type | Count |
|---|---|
| **Negative non-tumor** | 19,223 |
| **Negative tumor** | 26,308 |
| **Positive\* non-tumor** | 2,424 |
| **Positive tumor** | 9,433 |
| **Total** | **57,388** |

*Any staining pattern and intensity were considered as positive in Ki67.

Recommendations on semantic annotation for computational pathology

**Table S12** Usage of annotation constructs (rectangles) for cell and region annotation

| Construct | Count |
| --- | --- |
| **Box_All_Cell_20x** | 281 |
| **Box_All_Region_5x** | 80 |
| **Box_ALL_Special_20x** | 288 |
| **Box_ALL_Special_5x** | 67 |
| **Box_Cell_20x** | 1,792 |
| **Box_Region_5x** | 1,485 |
| **Box_Special_20x** | 3,518 |
| **Box_Special_5x** | 3,220 |
| **Total** | **10,731** |

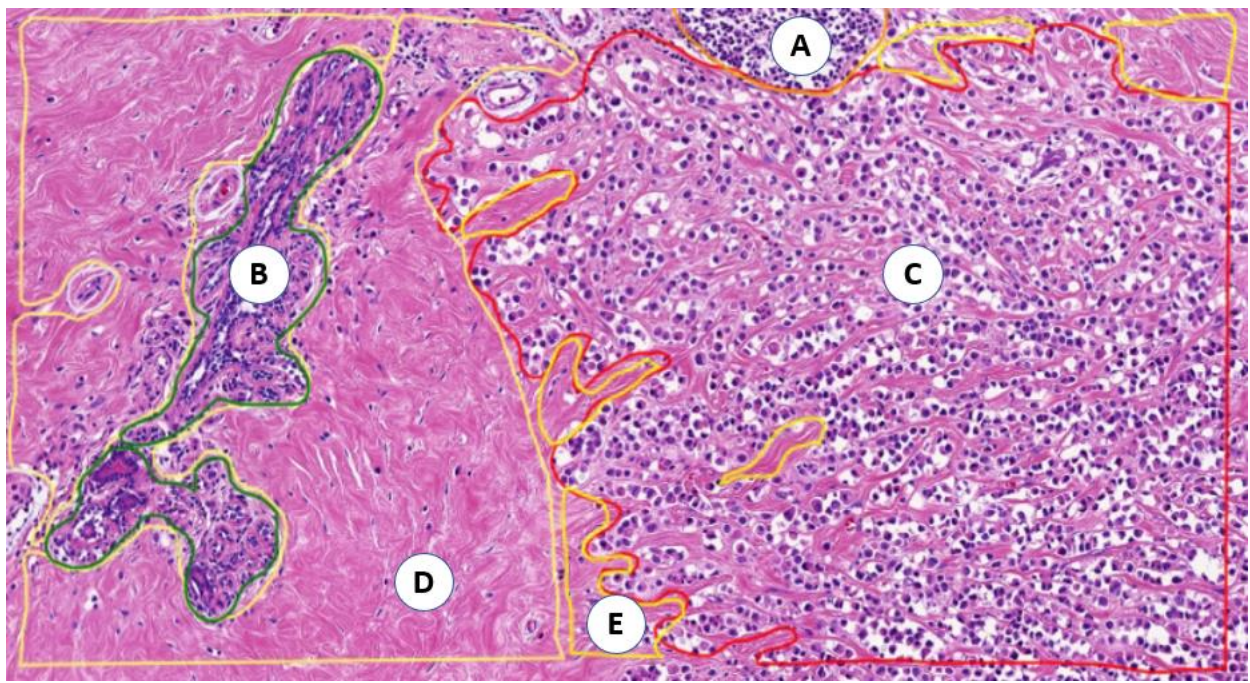

**Fig. S2** Region-level annotations. A) lymphoid aggregate, B) Normal terminal ductal lobular unit (TDLU), C) tumor, D) Stromal Fibroblast, E) Tumor-associated stroma

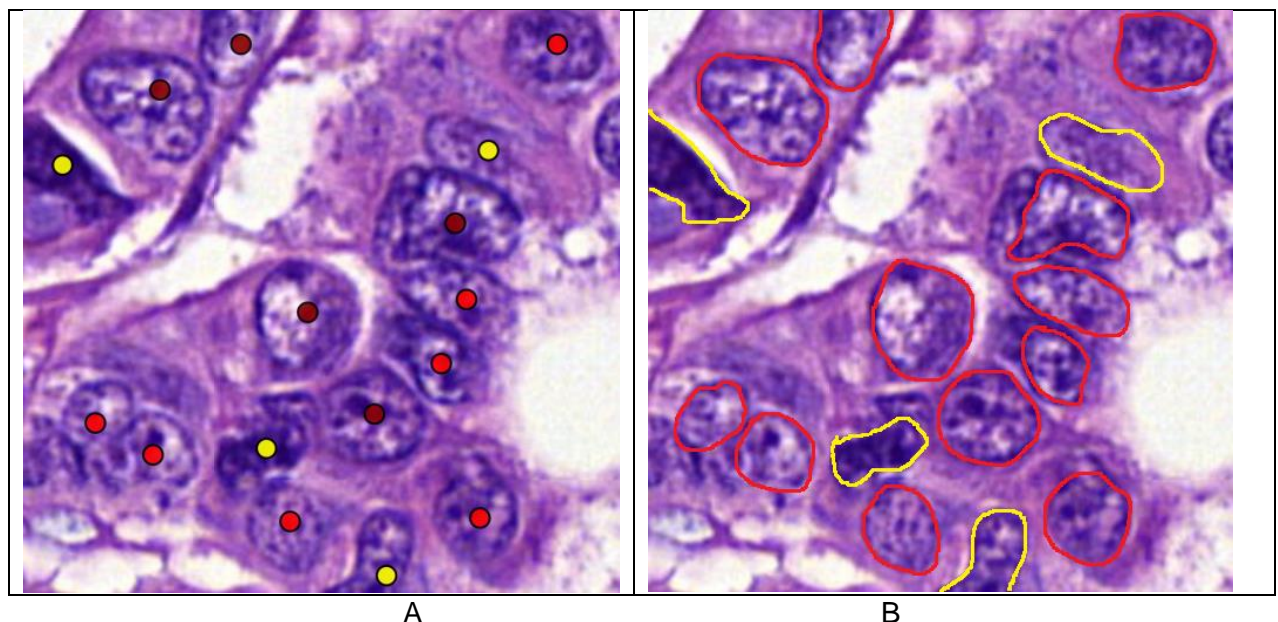

**Fig. S3** A) Point-based cell annotation, B) Free-hand polygon-based cell annotation. Different colors indicate different cell classes.

## S1.4 Supporting information on annotation constructs

**Bounding box:** Bounding boxes are axis-aligned rectangles which can be used for enclosing ROIs, or for enclosing other constructs such as polygons, lines, and points. Bounding boxes may be used for exhaustive annotations as it is almost impossible to annotate the WSI completely. A 'consensus box' can be drawn to allow multiple pathologists to annotate the same ROI. This box can be analyzed to assess concordance between experts when assessing key features within a WSI. For example, if a box is defined for region-level annotation, then the annotators can concentrate fully on this area while annotating. This also makes it easier for automation scripts to extract relevant annotation for ML purpose. Boxes can be further categorized, such as a region-box, cell-box, individual-annotation-box, consensus-annotation-box, etc.

**Point/circle/dot:** A point annotation is used to annotate cells or regions. This type of annotation is quicker but less precise as it will only provide a nominal center. If a more detailed annotation is required, such as nuclei or cell boundary, then a polygon or free-hand drawing is recommended (Fig. S4).

**Polygon:** A polygon can be used to provide additional precision to a region/cell annotation.

**Line:** A straight line can be drawn to segregate two neighboring structures, join two structures of interest, mark axis (e.g., major-axis or minor-axis) of a region/gland or cell, measure the diameter of an object or measure distance between two objects.

**Text:** Textual annotation can be used to describe regions and cells in the surrounding context. This annotation acts as a meta-annotation for other annotations.

For uniform annotations, we recommend that the workflow of the actual annotation process including launching the software, logging in, selection of images and selection of annotation tool should also be documented as a flowchart. A sample flowchart for region-level annotation is shown in Fig. S7.

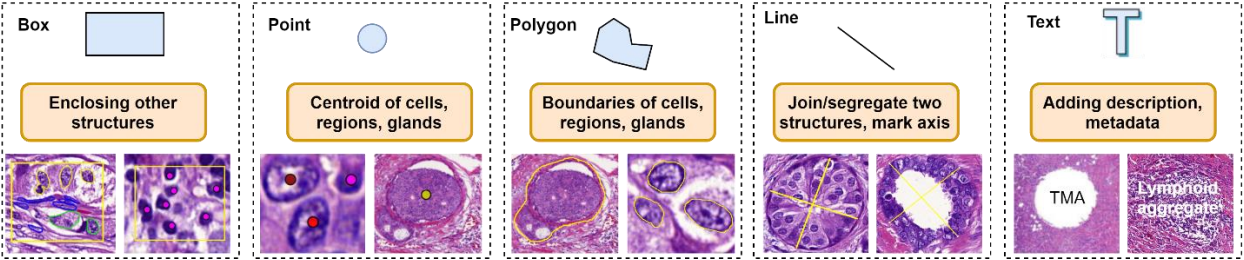

**Fig. S4** Different tools that can be used for annotations.

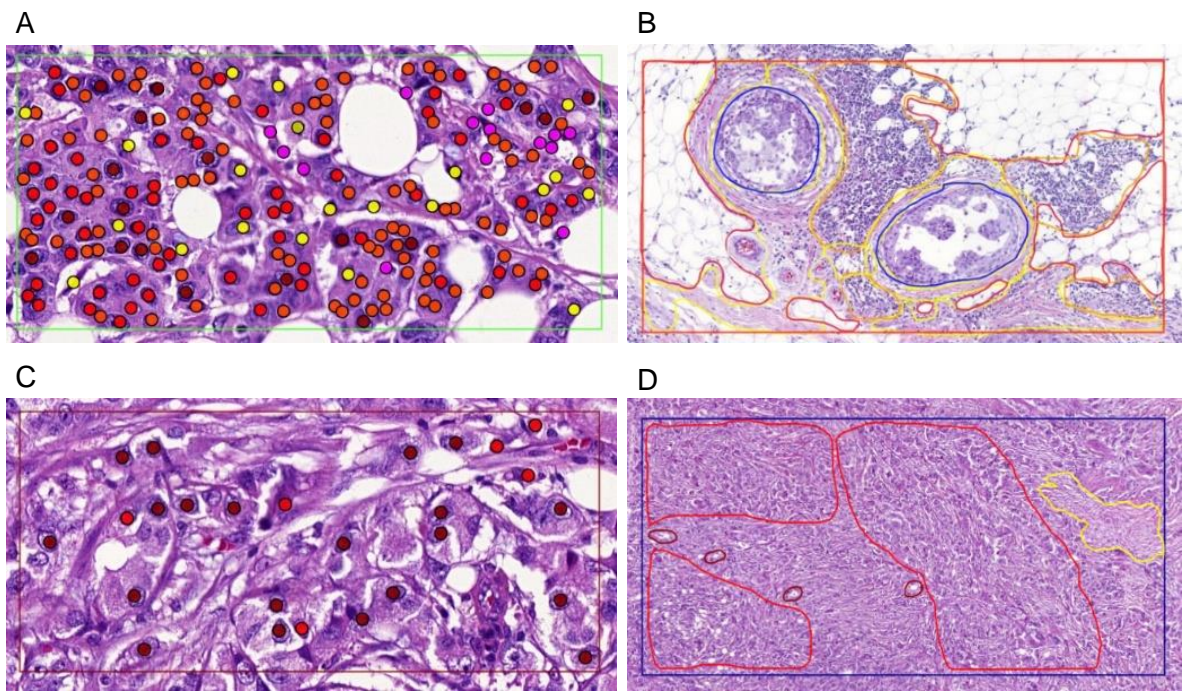

**Fig. S5** Exhaustive (A, B) vs non-exhaustive (C, D) annotations. A & B) Exhaustive cell-level and region-level annotations where all the cells/regions in the bounding box are annotated. C & D) Non-exhaustive cell-level and region-level annotations where some of the cells/regions in the bounding box are not annotated.

Recommendations on semantic annotation for computational pathology

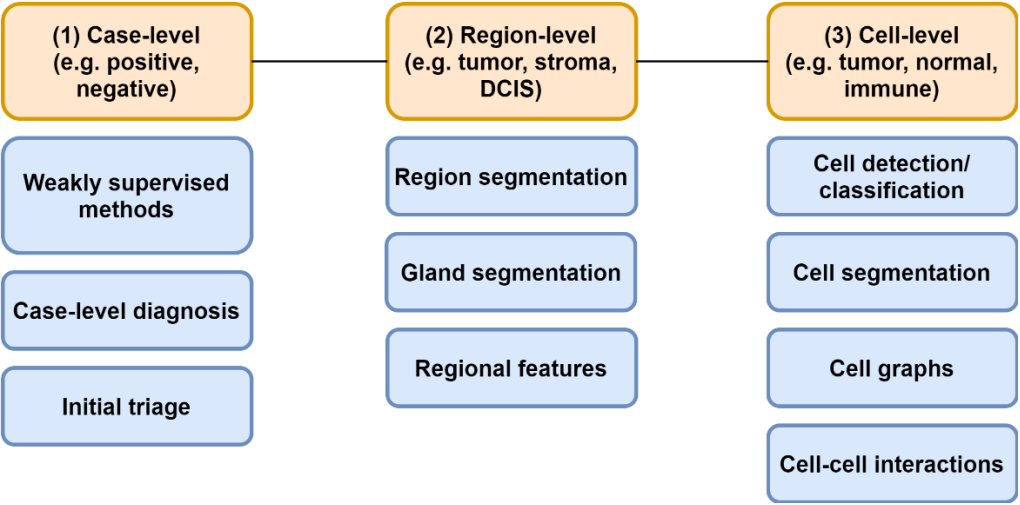

**Fig. S6** Proposed phases of annotations with their corresponding machine learning usage.

**Flowchart for H&E region-level annotation**

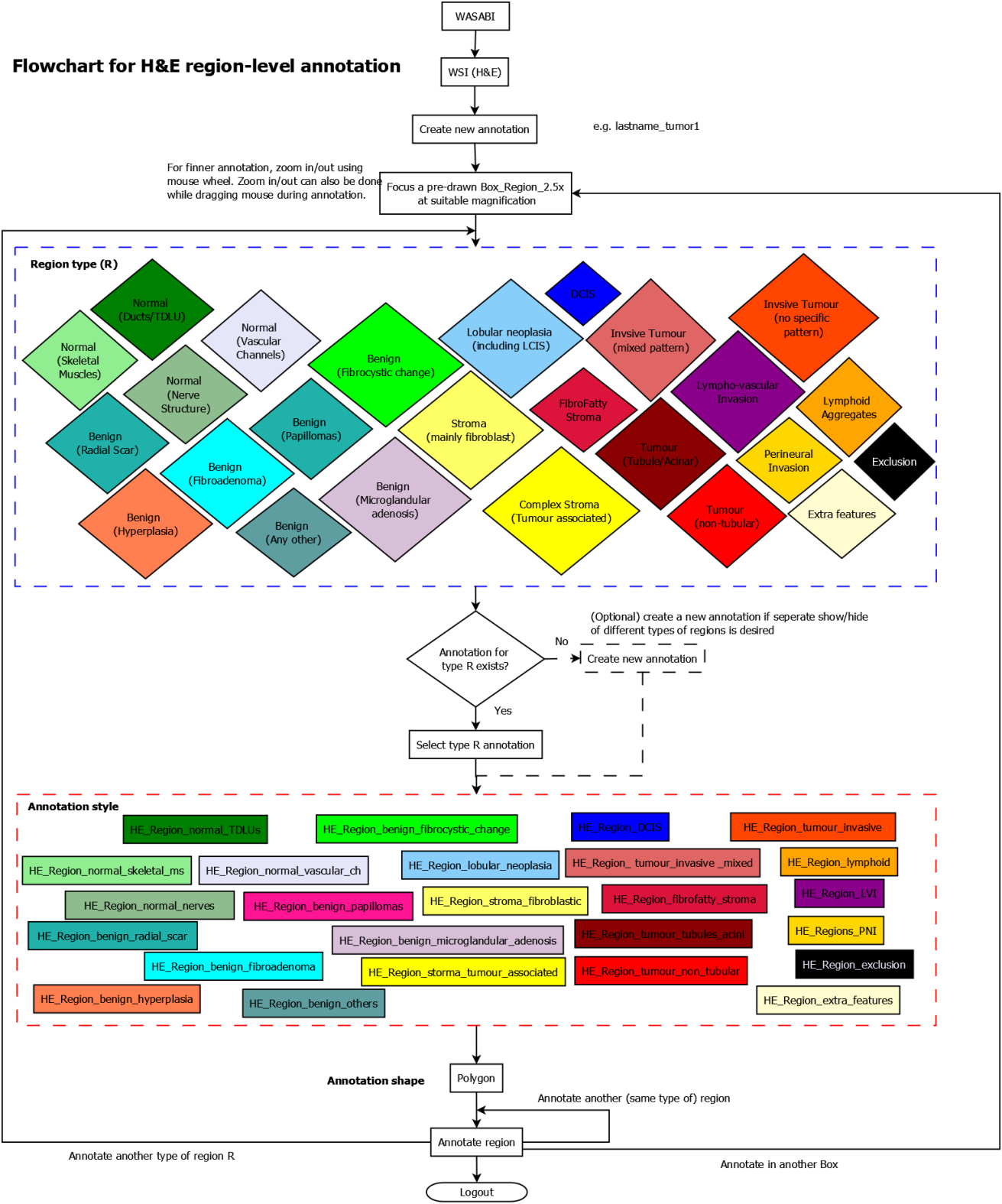

**Fig. S7** Flowchart for breast H&E region-level annotations

## S1.5 Supporting information on interactive annotations

For the segmentation task (requiring dense annotation), interactive approaches bring several advantages over fully automated approaches: 1) the supervisory signal acts as a prior to the interactive model, and leads to better performance; 2) interactive models are less sensitive to domain shift and are more generalizable (22); and 3) interactive models give the flexibility to the user to choose arbitrary instances of objects present in the visual field – for instance, selecting and focusing on only some of the nuclei to be segmented, instead of segmenting all of them using an automatic method.

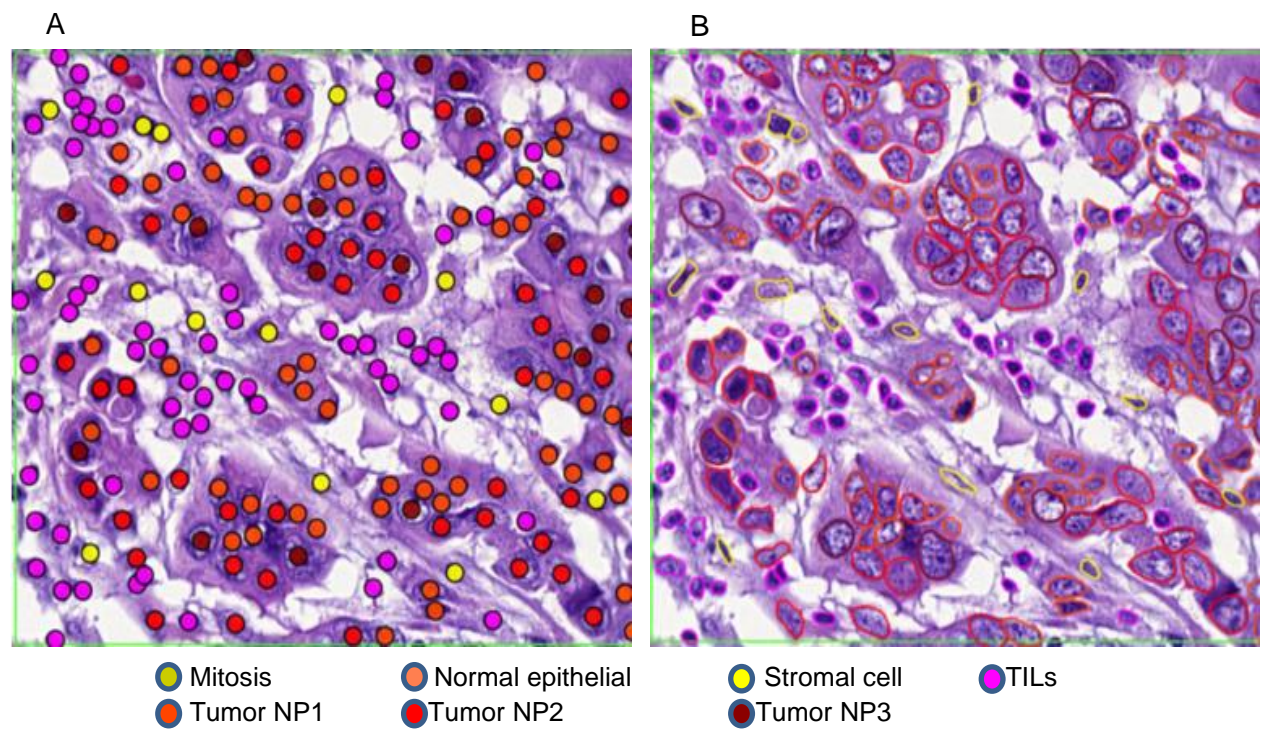

**Fig. S8** NuClick interactive segmentation. A) Point annotation for H&E cells, B) NuClick generated segmentation masks.

## S1.6 Supporting information on quality of tissue sectioning, staining and scanning

### S1.6.1 Tissue sectioning quality

For accurate and reliable quantitative analysis, enough tissue should be present on the slide. To avoid any false signal during ML based analysis, the section should be checked for any damage such as tears, folds, creases, and knife marks. The presence of artefacts such as blurriness, pen-markings, and tissue folding could pose major challenges to the downstream analyses. They are normally unintentionally introduced during both routine slide preparation and digitization. These kinds of artefacts can be manually excluded during annotation and review. However, manual review is time consuming, laborious, and subject to intra- and inter-reader variability therefore an automated QC of images is required. Slides which consist of large artefacts regions should be excluded directly before further annotation and computational analysis.

### S1.6.2 Staining quality

Proper SOPs should be followed for staining the slides (37,38). Staining issues could hamper initial analytical tasks such as nuclei detection and segmentation and any further downstream analyses relying on these steps. If the slides is good enough for diagnosis it is good enough for ML. Staining intensity should be harmonized by stain deconvolution steps to avoid problems of differences in performance by different scanners and laboratories.

### S1.6.3 Scanning quality

Though each digital whole slide scanner may have its own requirements and workflow, certain standards should be followed while scanning glass slides (39). The images should be checked for scanner resolution, out-of-focus, scan line and background artefacts.

## S2. Results (Applied example)

## S2.1 Quality control

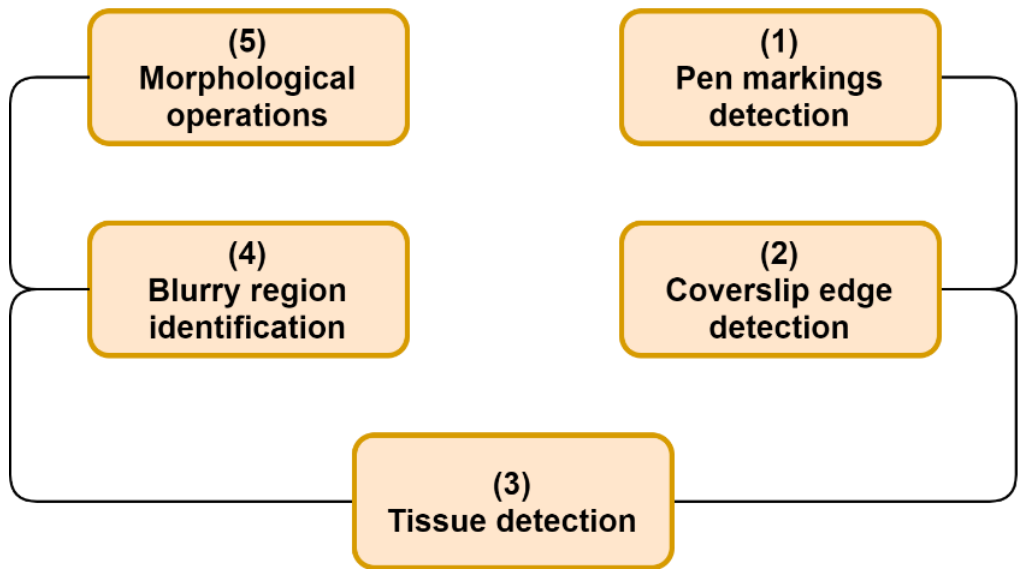

**Fig. S9** Workflow of ImageQC

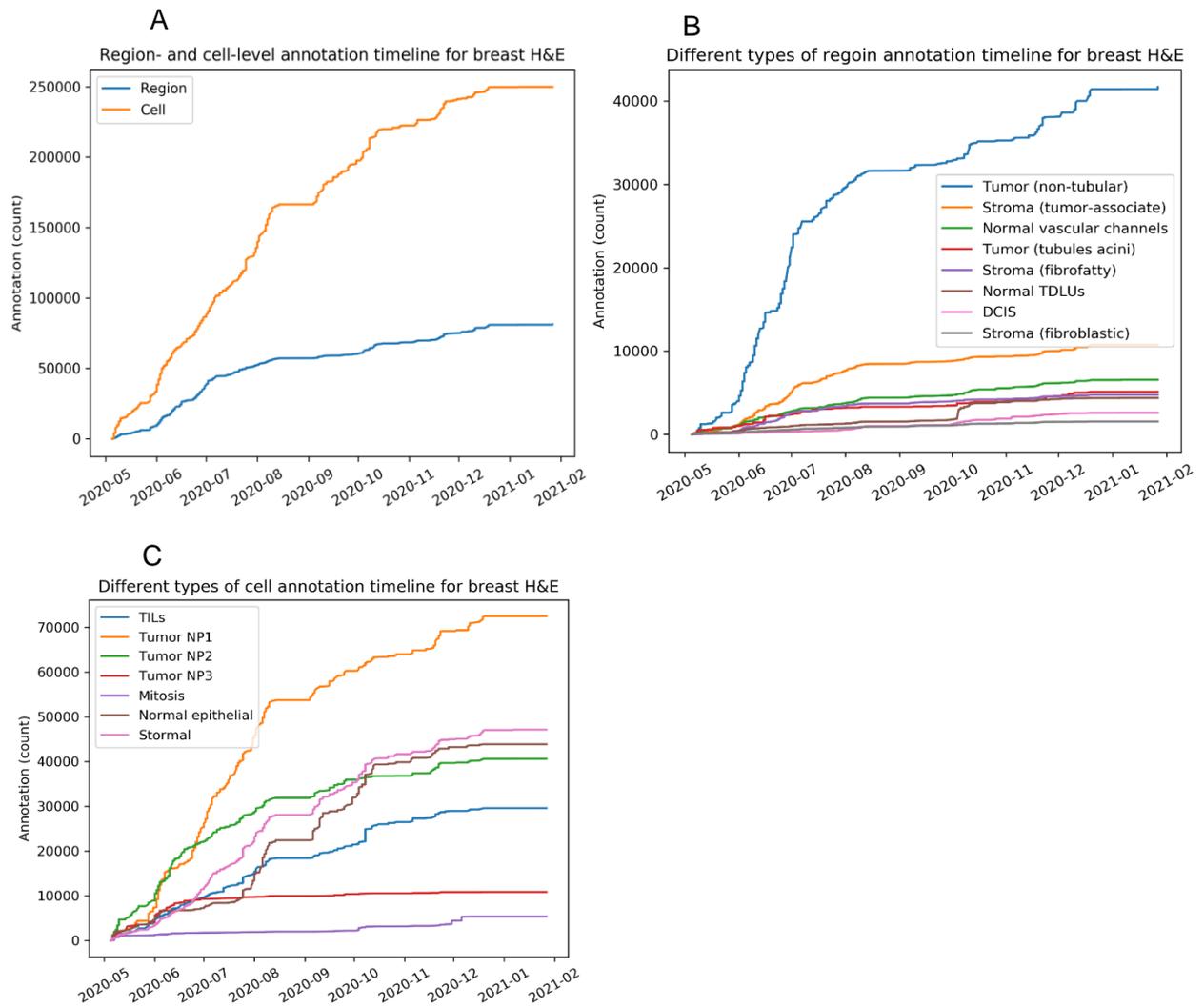

**Fig. S10** A) Timeline of overall region- and cell-level annotation for breast H&E showing increase in the counts of annotations over time; B) Timeline of different types of breast H&E regions (top 8); C) Timeline of different types of breast H&E cells. NP (nuclear pleomorphism)

## S2.2 Annotation concordance and usage

**Table S13** Confusion matrix of different types of breast H&E cell annotation by two pathologists

| | Cell type | Pathologist 2 | | | | | | |
|---|---|---|---|---|---|---|---|---|
| | | Mitosis | Normal epithelial | Stroma | TILs | Tumor NP1 | Tumor NP2 | Tumor NP3 |
| **Pathologist 1** | Mitosis | **43** | 0 | 2 | 0 | 6 | 0 | 0 |
| | Normal epithelial | 0 | **318** | 61 | 0 | 1 | 146 | 5 |
| | Stroma | 0 | 0 | **2,852** | 0 | 530 | 397 | 9 |
| | TILs | 1 | 26 | 469 | **1,600** | 162 | 38 | 5 |
| | Tumor NP1 | 0 | 0 | 0 | 0 | **4,360** | 4 699 | 352 |
| | Tumor NP2 | 0 | 0 | 0 | 0 | 0 | **1,353** | 638 |
| | Tumor NP3 | 0 | 0 | 0 | 0 | 0 | 0 | **207** |

**Table S14** Using cell-level annotation for machine learning. Pr (precision), Re (recall)

| Model | Overall (Macro-avg) | | | Tumor | | | TILs | | | Stromal | | | Norm-epithelial | | |
|---|---|---|---|---|---|---|---|---|---|---|---|---|---|---|---|
| | Pr | Re | F1 | Pr | Re | F1 | Pr | Re | F1 | Pr | Re | F1 | Pr | Re | F1 |
| *M1 | 0.68 | 0.49 | **0.53** | 0.79 | 0.97 | **0.87** | 0.95 | 0.62 | **0.76** | 0.88 | 0.29 | **0.43** | 0.08 | 0.07 | **0.07** |
| *M2 | 0.78 | 0.80 | **0.79** | 0.94 | 0.92 | **0.93** | 0.92 | 0.87 | **0.90** | 0.84 | 0.80 | **0.82** | 0.42 | 0.61 | **0.50** |

*M1: Pretrained HoVer-Net

*M2: Fine-tuned HoVer-Net

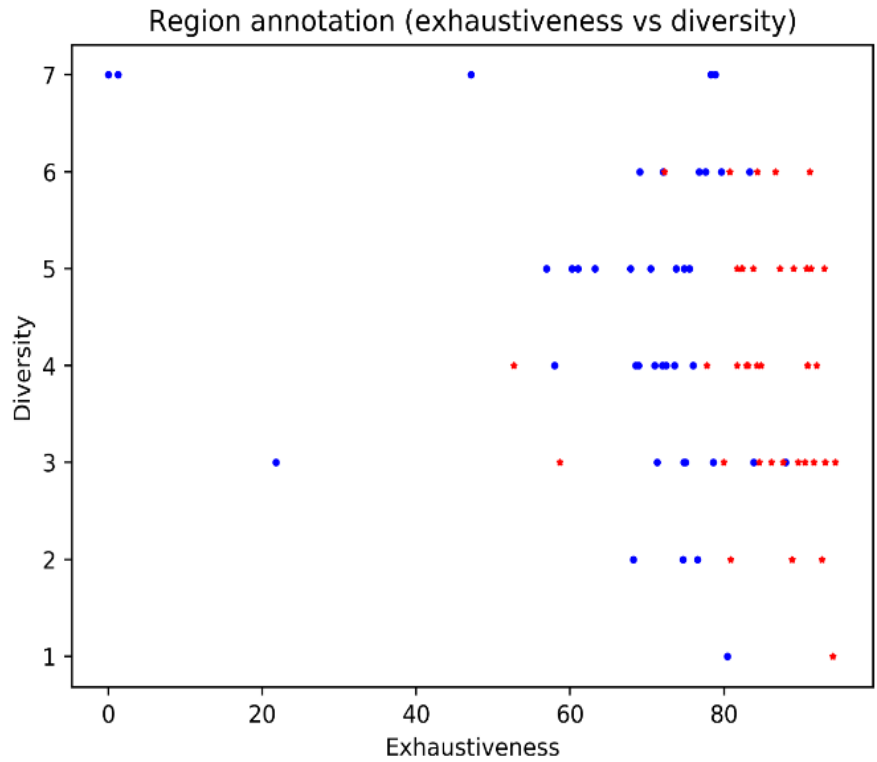

**Fig. S11** Exhaustiveness vs Diversity of the region box annotation where each dot represents a region box. Blue dots represent consensus boxes whereas red dots represent individual boxes.

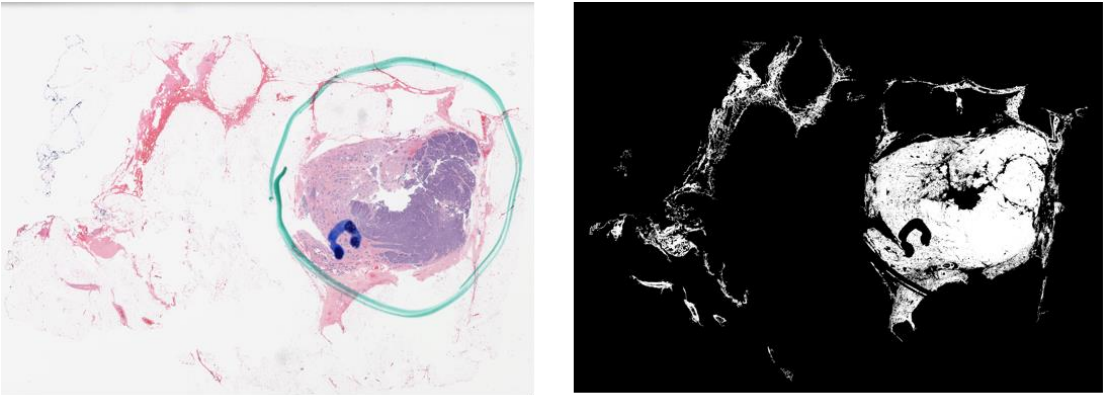

(a)

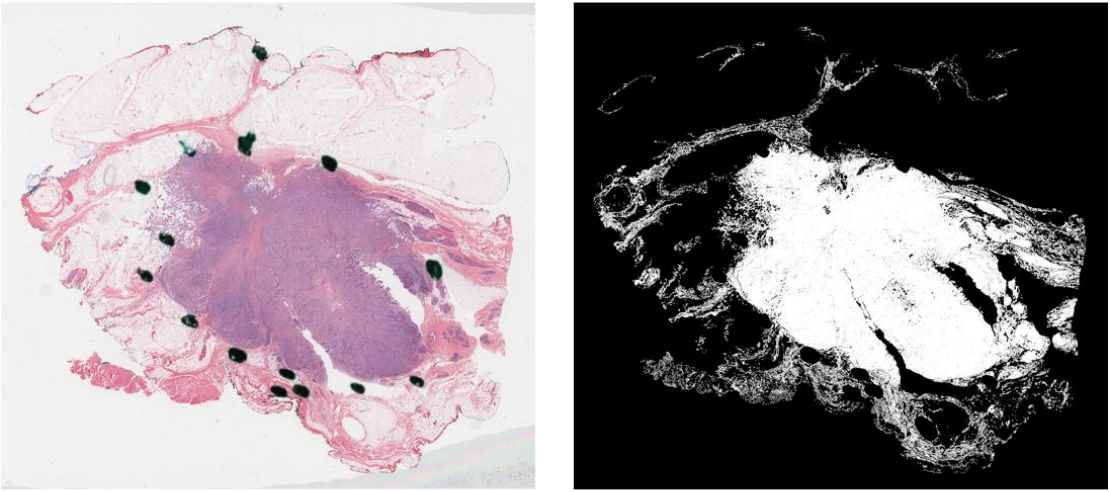

(b)

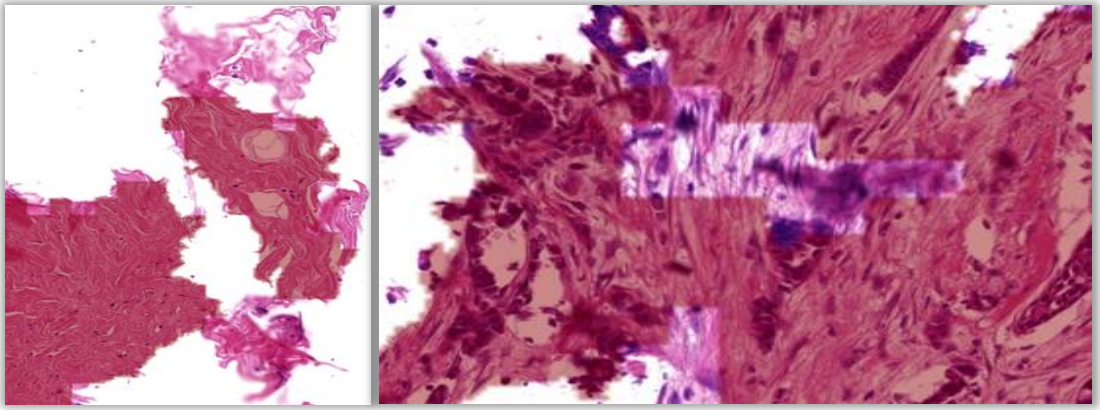

(c)

**Fig. S12** Artefact detection during ImageQC. (a) and (b) showing pen marking on a WSIs and its artefact-free mask. (c) showing artefact-free area in red shade, non-shaded regions showing blurriness and pen marking.

# Recommendations on semantic annotation for computational pathology

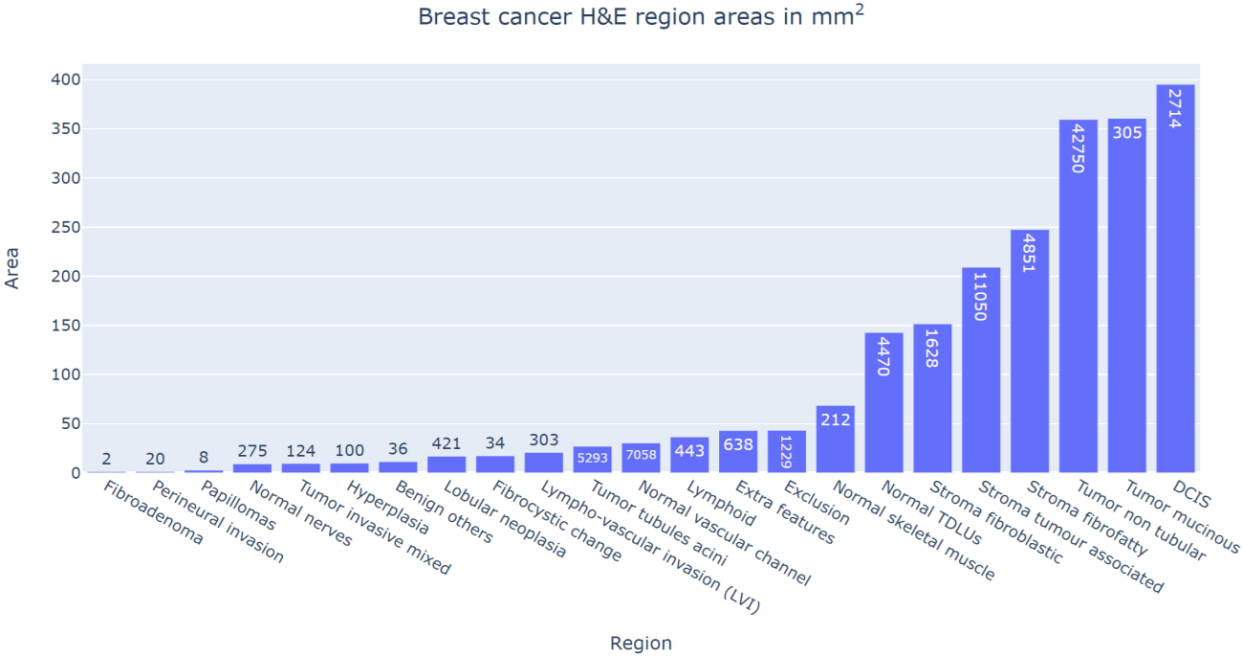

**Fig. S13** Breast H&E region annotation count and area for 314 cases. The numbers on the bars denote the number of regions annotated whereas the numbers on the y-axis represent the area in mm².

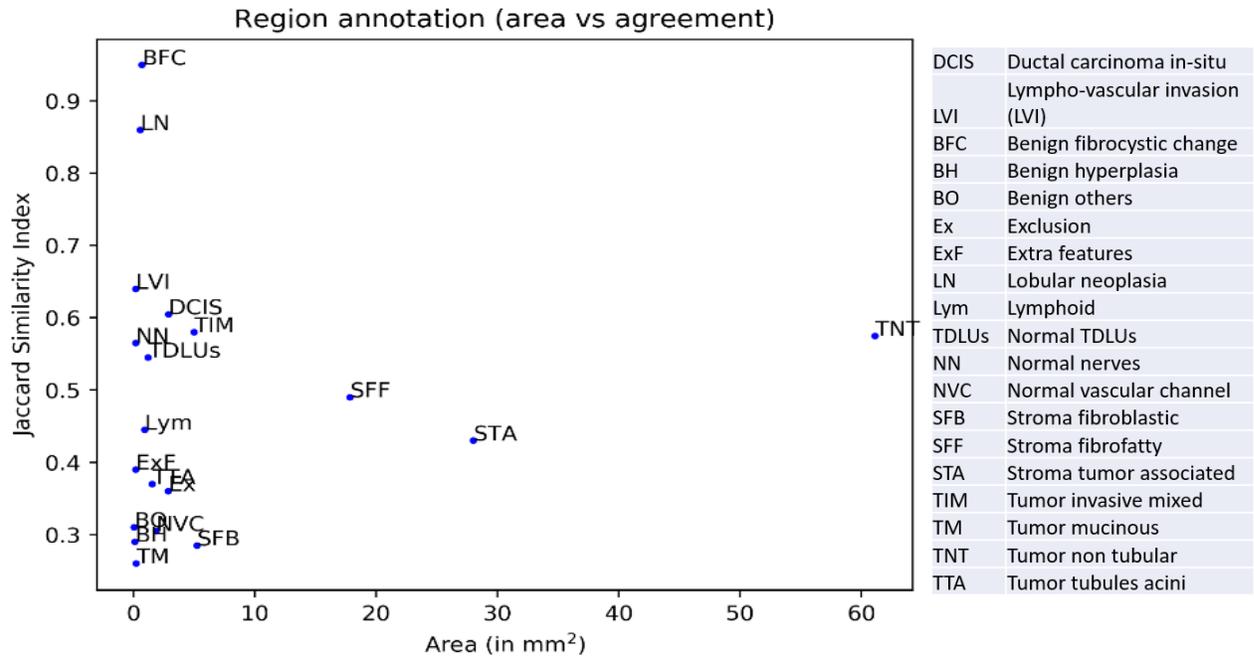

**Fig. S14** Inter-pathologist agreement on sample regions of breast H&E images. x-axis shows the area in mm$^2$ whereas y-axis shows Jaccard Similarity Index.